\documentclass[prd,preprintnumbers,nofootinbib]{revtex4} 
\usepackage{graphicx} 
\usepackage{amsmath}
\usepackage{amsfonts,amsbsy}
\usepackage{amssymb}
\usepackage{appendix}
\usepackage{slashed}

\def\be{\begin{equation}}
\def\ee{\end{equation}}
\def\bea{\begin{eqnarray}}
\def\eea{\end{eqnarray}}

\def\eq#1{{Eq.\,(\ref{#1})}}
\def\fig#1{{Fig.~\ref{#1}}}
\def\fig#1{{Fig.~\ref{#1}}}

\def\k{{\mathbf k}}

\def\p{{\mathbf p}}

\def\q{{\mathbf q}}
\def\k{{\mathbf k}}

\def\x{{\mathbf x}}
\def\y{{\mathbf y}}

\begin{document}

\title{\bf  Multi Quark Production in p+A collisions: Quantum Interference Effects}

\preprint{CERN-TH-2018-007}

\author{Alex Kovner$^{1,2,3,4,5}$ and Amir H. Rezaeian$^{2,3}$}

\affiliation{
$^1$ Dept. of Physics, University of Connecticut, Storrs, CT 06269, USA\\
$^2$ Departamento de F\'\i sica, Universidad T\'ecnica
Federico Santa Mar\'\i a, Avda. Espa\~na 1680,
Casilla 110-V, Valparaiso, Chile\\
$^3$  Centro Cient\'\i fico Tecnol\'ogico de Valpara\'\i so (CCTVal), Universidad T\'ecnica
Federico Santa Mar\'\i a, Casilla 110-V, Valpara\'\i so, Chile\\
$^4$ Physics Department, Ben Gurion University of the Negev, Beer Sheva, Israel  \\
$^5$Theoretical Physics Department, CERN, CH-1211 Geneve 23, Switzerland\\
}

\begin{abstract}
We consider forward inclusive production of several quarks in the  high energy  p-A collisions in the CGC formalism. 
For three particle production we provide a complete expression in terms of multipole scattering amplitudes on the nucleus and multi particle generalized TMD's of the proton. We then calculate all the terms that are not suppressed by the factor of the area in four particle production, and generalize this result up to terms of order $1/N_c^2$ for arbitrary number of produced particles. Our results include the contribution of quantum interference effects both in the final state radiation (HBT) and in the initial projectile wave function (Pauli blocking).
\end{abstract}

\maketitle

\section{Introduction}
The observation of ridge correlations  in p-p and p-Pb collisions at LHC \cite{exp-r1,exp-r2,exp-r3,exp-r4,exp-r5} provided strong impetus for study of correlations in particle production at high energy. Two main physical sources  of such correlations have been advocated in recent years:  strong collective effects due to final state interactions \cite{hydro}, and initial state effects due to ``quasi collectivity'' - correlations inherited from the nontrivial correlated structure of the initial state \cite{ridge0, ridge1,ridge2,ridge3,ridge1-raju,ridge33,ridge4,ridge5}. The origin of the initial state induced correlations has been better understood in the last couple of years. In addition to the ``classical scattering effects'' which collimate the emitted particles that scatter off correlated adjacent regions of the target \cite{ridge2,ridge3,ridge5}, an important (and in some circumstances leading) role is played by quantum interference effects. These last come in two varieties: the variant of {\bf partonic} Hanbury-Brown, Twiss  correlations \cite{hbthadrons} and quantum statistics effects in the incoming projectile wave function \cite{bosee,paulib}. 

In our previous work \cite{2pc} we have considered in detail inclusive  production of two quarks, whether identical or non-identical. We have explicitly identified  the quantum interference contributions, and have shown that they produce a parametrically leading effect in production of fundamental quarks. We have also argued that this effect in gluon production is of the same order as other effects studied so far, and therefore cannot be neglected. The aim of the present paper is to extend this analysis to inclusive production of more than two fundamentally charged particles. 
The framework of our approach is identical to that in \cite{2pc}. We use  the extension of the hybrid formalism to include the multiple-parton-scattering (MPS) in the Color-Glass-Condensate (CGC) approach \cite{kr1}.  
We will be working within a variant of the "hybrid" approximation \cite{dj-rhic} which is appropriate for forward particle production. In the hybrid  CGC approach, we assume that the small-x gluon modes of the nucleus have a large occupation number so that the target nucleus can be described in terms of a classical color field.  This should be a good approximation for large enough nucleus at high-energy. This color field emerges from the classical Yang-Mills equation with a source term provided by faster partons. The renormalization group equations which govern the separation between the soft and hard models are then given by the non-linear Jalilian-Marian, Iancu, McLerran, Weigert, Leonidov, Kovner (JIMWLK) \cite{jimwlk} and Balitsky-Kovchegov (BK) \cite{bk} evolution equations.  We further assume that the projectile proton is in the dilute regime and can be described in ordinary perturbative approach using parton picture like assumptions. This somewhat restricts the validity range of our approximation as discussed in \cite{2pc}.

We also neglect processes where the quarks are produced from the splitting of scattered projectile gluons \cite{private}. For $n$ quark production this approximation is parametrically leading for projectiles which contain at least $n$ valence quarks. Thus for $n>3$ our calculation is more appropriate for a light nucleus projectile rather than a proton in the sense that for a proton projectile our results may get significant corrections. These corrections can be in principle straightforwardly calculated in perturbation theory, but we do not endeavor this calculation in the present paper.

The plan of the paper is as follows. In Section II we calculate in detail production of three quarks following  the method of \cite{2pc} and exhibit all the different contribution to the production: the ones due to quantum interference effects, which are not suppressed by  a power of area, as well as the aforementioned ``classical'' effects which have an area suppression. In Section III we  extend the calculation to four quarks. This time however we consider only the quantum interference terms, as complete calculation becomes rather long and cumbersome. In Section IV we generalize the results to arbitrary number of produced particles. Here we only consider the quantum interference terms and only contributions to production cross section down to order $1/N_c^2$. In principle the analysis can be extended to higher orders in $1/N_c$ as well, but the expressions become rather lengthy and we refrain from recording them here. We close by a short discussion in Section V.

We note two interesting recent papers \cite{vladi} and \cite{bmu} which consider the quantum interference leading to the  HBT effect in gluon production, albeit from a somewhat different vantage point. Compared to these works the problem of emission of fundamental charges  is calculationally simpler and thereby also admits a more controlled treatment. Hopefully the two approaches will converge in near future and we can have a more complete picture of the importance of quantum interference effects in particle production, and whether they are relevant for explanation of the experimental data.

\section{Inclusive three quark production in proton-nucleus collisions}
We start our analysis by considering the  inclusive production of three quarks. 
\begin{figure}[t]                                       
                                  \includegraphics[width=9 cm] {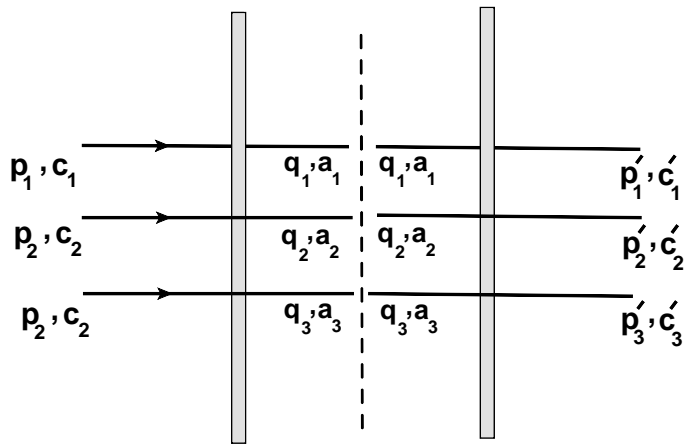}             
\caption{The diagram contributing to three quark production  in the background of the CGC field. The diagrams on the left and right side of the dashed line correspond to the amplitude and the complex conjugate amplitude.  The shaded box (the CGC shock waive) denotes the interaction of a quark to all orders with the background field via multiple gluon exchanges.  The color ($c_i, c'_i$), spin ($s_i, s'_i$) and momenta ($p_i, p'_i$) of the $i$-th quark in the amplitude and the complex conjugate amplitude  are also shown.  }
\label{f2}
\end{figure}

The cross-section for production of three quarks with momentum $q_1, q_2, q_3$ in the proton-nucleus (p-A) scatterings  can be written in the following general form, 
\be d\, \sigma^{p+A\to qqq+X} =  \frac{d^3 q_1}{(2\pi)^3\, 2
  q_1^-} \frac{d^3 q_2}{(2\pi)^3\, 2
  q_2^-} \frac{d^3 q_3}{(2\pi)^3\, 2
  q_3^-}
  \ \langle| \langle \text{jet}(q_1), \text{jet}(q_2)\text{jet}(q_3)| \text{Proton} \rangle|^2 \rangle_{\text{color sources}}, 
\label{m1}
\ee
where $|\text{Proton}\rangle$ is the wave function of the energetic  proton with vanishing transverse momentum, and the averaging should be performed over the distribution of the color charges in the target.

The wave function of the proton can be written generally as \cite{kr1}, 
\begin{equation}
|\text{Proton} \rangle=\sum_X\sum_{c_1,s_1}\sum_{c_2,s_2}\sum_{c_3,s_3}\int\int\frac{d^3p_1}{(2\pi)^3}\frac{d^3p_2}{(2\pi)^3} \frac{d^3p_3}{(2\pi)^3}\tilde A(p_1,c_1,s_1;p_2,c_2,s_2;p_3,c_3,s_3;X)|p_1,c_1,s_1;p_2,c_2,s_2;p_3,c_3,s_3;X\rangle, 
\end{equation}
where $(p_i,c_i,s_i)$ labels the momentum, color and spin of the $i$-th quark (see \fig{f2}), and $X$ labels the configuration of all the spectator partons in the proton. 

The S-matrix element of the proton scattering into the state with three quarks and an arbitrary configuration of spectator particles can be written as
\begin{eqnarray}
\langle q_1,a_1;q_2,a_2;q_3,a_3;X'|\text{Proton} \rangle&=&\sum_X\sum_{c_1,c_2,c_3}\int\int\int \frac{d^3p_1}{(2\pi)^3}\frac{d^3p_2}{(2\pi)^3}\frac{d^3p_3}{(2\pi)^3} \nonumber\\
&\times&\tilde A(p_1,c_1;p_2,c_2;p_3,c_3;X)\langle q_1,a_1; q_2,a_2;q_3,a_3;X'|p_1,c_1;p_2,c_2; p_3,c_3;X\rangle.  \
\end{eqnarray}
In the above $\langle A|B\rangle$ denotes the S-matrix element of the initial state $B$ scattering  into the final state $A$. In the spirit of the parton model we assume that partons scatter independently. For three distinct quarks this translates into
\begin{equation}
\langle q_1,a_1;q_2,a_2;q_3,a_3,X'|p_1,c_1;p_2,c_2;p_3,c_3, X\rangle=\langle q_1,a_1|p_1,c_1\rangle \langle q_2,a_2|p_2,c_2\rangle \langle q_3,a_3|p_3,c_3\rangle\langle X'|X\rangle. 
\end{equation}
With the above definitions we can write down the expression for the triple  inclusive cross-section  as
\begin{eqnarray}\label{cross}
&&\mathcal{I}=\sum_{X'} |\langle q_1,a_1;q_2,a_2;q_3,a_3;X'|\text{Proton}\rangle|^2=\sum_{c_1,c_2,c_3,c'_1,c'_2,c'_3,a_1,a_2,a_3}\int_{p_1,p_2p_3,p'_1,p'_2,p'_3}\sum_X \nonumber\\ &\times& A(p_1,c_1;p_2,c_2;p_3,c_3;X)A^*(p'_1,c'_1;p'_2,c'_2; p'_3,c'_3, X)\nonumber\\
&\times& \Bigg[\langle q_1,a_1|p_1,c_1\rangle \langle q_1,a_1|p'_1,c'_1\rangle^* \Bigg]
\Bigg[\langle q_2,a_2|p_2,c_2\rangle
 \langle q_2,a_2|p'_2,c'_2\rangle^*\Bigg]
 \Bigg[\langle q_3,a_3|p_3,c_3\rangle
 \langle q_3,a_3|p'_3,c'_3\rangle^*\Bigg], \
\end{eqnarray}
where  for distinct quarks we have $A\equiv \tilde  A$, while for the identical quark case, the amplitude $ A$ is the completely antisymmetric part of  amplitude  $\tilde A$, where the anti symmetrization is performed  with respect to momenta, spin and color of the three quarks. 

The above expression is only valid under the parton model assumption, namely for the cases that the typical transverse momentum of the quarks in the proton wave function is much smaller than the momentum of the produced particles. If this is not the case, additional terms arise in the expression for the cross-section which involve scattering of the ``spectator'' particles. The evaluation of these extra terms requires the knowledge of complicated matrix elements, which goes beyond our present ability, see discussion in \cite{2pc}. Throughout this paper we therefore limit ourselves to consideration of large transverse momentum of produced quarks.

For the single quark scattering amplitude we have \cite{gf}, 
\begin{equation}\label{ampl}
\langle q_1,a_1|p_1,c_1\rangle=2\pi\delta(p_1^{-}-q_1^-)\frac{1}{\sqrt{2p^-_1}} \int d^2\x e^{i(\p_1-\q_1)\x}[U(\x)]_{a_1c_1}\bar u(q)\gamma^-u(p_1), 
\end{equation}
where $U(\x)$ is the scattering matrix of a quark on the colored glass condensate target, and it is represented as a unitary matrix in fundamental representation of $SU(N_c)$. 
The factor $\frac{1}{\sqrt{2p^-_1}}$ was introduced for convenience in order to avoid extra normalization factor in the cross-section defined in \eq{m1}.  
Throughout the paper we denote transverse coordinates and momenta by boldface letters.  In the following we use the standard relation between spinors, namely $\bar u_s(q)u_{s'}(q)=\slashed{q}_{ss'}$. Using Eq.\,(\ref{ampl}), the cross-section is written as
\begin{eqnarray}
\mathcal{I}&=&(2\pi)^6\delta(p_1^{-}-q_1^-)\delta(p_1^{\prime -}-q_1^-)\delta(p_2^--q_2^{-})\delta(p_2^{\prime -}-q_2^{-})
\delta(p_3^{-}-q_3^-)\delta(p_3^{\prime -}-q_3^-)
\frac{1}{8q_1q_2q_3}\nonumber\\
&&\sum_{c_1,c_2,c_3, c'_1,c'_2,c'_3,a_1,a_2, a_3}\int_{\p_i,\p'_i,\x_i} \times\langle[U^\dagger(\x_1)]_{c'_1a_1}[U(\x'_1)]_{a_1c_1}[U^\dagger(\x_2)]_{c'_2a_2}[U(\x'_2)]_{a_2c_2} [U^\dagger(\x_3)]_{c'_3a_3}[U(\x'_3)]_{a_3c_3}\rangle \nonumber\\
&&e^{i[(\p'_1-\q_1)\x_1+(\q_1-\p_1)\x'_1+(\p'_2-\q_2)\x_2+(\q_2-\p_2)\x'_2] + (\p'_3-\q_3)\x_3+(\q_3-\p_3)\x'_3] }\nonumber\\
&&\bar u(p'_1)\gamma^-\slashed{q_1}\gamma^- u(p_1)\bar u(p'_2)\gamma^-\slashed{q_2}\gamma^-u(p_2)  \bar u(p'_3)\gamma^-\slashed{q_3}\gamma^- u(p_3) \nonumber\\
&&\sum_X A(p_1,c_1;p_2,c_2;p_3,c_3;X)A^*(p'_1,c'_1;p'_2,c'_2;p'_3,c'_3, X),  \label{m-2}\
\end{eqnarray}
where for brevity, we used the notation  $\int \frac{d^2\p}{(2\pi)^2}\equiv \int_p$. In the above all spin indexes and summation over spins are implicit. 

In the high energy limit one can perform the spin algebra in a straightforward way.
First,  we have
\begin{equation}
\gamma^-\slashed{q}\gamma^-= 2\gamma^-q^-. 
\end{equation}
In the  approximation where the largest component of momentum $p_i^\mu$ is $p_i^-$ (and the same for $p_i'$)
 the spinors do not depend on the transverse momentum, so that for different momenta  they only differ by a normalization factor 
$\frac{1}{\sqrt{p^-_1}}u^{s}(p_1)=\frac{1}{\sqrt{p^{'-}_1}}u^{s}(p'_1)$.  Therefore, at high energies we have
\begin{equation}
\bar u_{s'_1}(p'_1)\gamma^-\slashed{q}\gamma^- u_{s_1}(p_1)\bar u_{s'_2}(p'_2)\gamma^-\slashed{q}'\gamma^-u_{s_2}(p_2)=16 q^-q'^{-}\sqrt{p_1^-p_2^-p'^{-}_1 p^{\prime -}_2}
\delta_{s_1s'_1}\delta_{s_2s'_2}.
\end{equation}
Using the above relation, we can simplify the spin algebra in Eq.\,(\ref{m-2}) and obtain 
\begin{eqnarray}
\mathcal{I}&=&16(2\pi)^4  \delta(p_1^{-}-q_1^-)\delta(p_1^{\prime -}-q_1^-)\delta(p_2^--q_2^{-})\delta(p_2^{\prime -}-q_2^{-})
\delta(p_3^{-}-q_3^-)\delta(p_3^{\prime -}-q_3^-) q_1^-q_2^{-}q_3^{-} \nonumber\\
&&\sum_{c_1,c_2,c_3, c'_1,c'_2,c'_3,a_1,a_2, a_3}\int_{\p_i,\p'_i,\x_i} \times\langle[U^\dagger(\x_1)]_{c'_1a_1}[U(\x'_1)]_{a_1c_1}[U^\dagger(\x_2)]_{c'_2a_2}[U(\x'_2)]_{a_2c_2} [U^\dagger(\x_3)]_{c'_3a_3}[U(\x'_3)]_{a_3c_3}\rangle \nonumber\\
&&e^{i[(\p'_1-\q_1)\x_1+(\q_1-\p_1)\x'_1+(\p'_2-\q_2)\x_2+(\q_2-\p_2)\x'_2] + (\p'_3-\q_3)\x_3+(\q_3-\p_3)\x'_3] }\nonumber\\
&&\sum_X A(p_1,c_1;p_2,c_2;p_3,c_3;X)A^*(p'_1,c'_1;p'_2,c'_2;p'_3,c'_3, X). \
\end{eqnarray}
We have for now suppressed the spin dependence of the amplitude, but will deal with the question of polarization in the amplitude later on.

\subsection{The color algebra}
Averaging over the eikonal scattering matrices has to be performed in the target ensemble. We will use the fact that the target ensemble is globally color invariant. The average of any tensor  in such an ensemble has to be proportional to a linear combination  of available invariant tensors. Consequently for any such ensemble we must have
\begin{eqnarray}
&&\langle[U^\dagger(\x_1)]_{c'_1a_1}[U(\x'_1)]_{a_1c_1}[U^\dagger(\x_2)]_{c'_2a_2}[U(\x'_2)]_{a_2c_2} [U^\dagger(\x_3)]_{c'_3a_3}[U(\x'_3)]_{a_3c_3}\rangle\nonumber\\ 
&&=\delta_{c'_1c_1}[\mathcal{A}_1 \delta_{c'_2c_2}  \delta_{c'_3c_3} +\mathcal{A}_2 \delta_{c'_2c_3}\delta_{c_2c'_3}] 
+\delta_{c'_1c_2}[\mathcal{A}_3 \delta_{c'_2c_1}  \delta_{c'_3c_3} +\mathcal{A}_4 \delta_{c'_2c_3}\delta_{c_1c'_3}]  
+\delta_{c'_1c_3}[\mathcal{A}_5 \delta_{c'_2c_2}  \delta_{c'_3c_1} +\mathcal{A}_6 \delta_{c'_2c_1}\delta_{c_2c'_3}]. \
\end{eqnarray}
Tracing  this relation over different pairs of indexes we obtain a simple set of linear equations for $\mathcal{A}_i$
\begin{eqnarray}
&&N_c^3\mathcal{A}_1+N_c^2\mathcal{A}_2+N_c^2\mathcal{A}_3+N_c\mathcal{A}_4+N_c^2\mathcal{A}_5+N_c\mathcal{A}_6=N_c^3\langle D(\x_1,\x'_1)D(\x_2,\x'_2)D(\x_3,\x'_3\rangle),\nonumber\\
&&N_c^2\mathcal{A}_1+N_c\mathcal{A}_2+N_c^3\mathcal{A}_3+N_c^2\mathcal{A}_4+N_c\mathcal{A}_5+N_c^2\mathcal{A}_6=N_c^2\langle Q(\x_1,\x'_1,\x_2,\x'_2)D(\x_3,\x'_3)\rangle,\nonumber\\
&&N_c^2\mathcal{A}_1+N_c^3\mathcal{A}_2+N_c\mathcal{A}_3+N_c^2\mathcal{A}_4+N_c\mathcal{A}_5+N_c^2\mathcal{A}_6=N_c^2\langle Q(\x_3,\x'_3,\x_2,\x'_2)D(\x_1,\x'_1)\rangle,\nonumber\\
&&N_c^2\mathcal{A}_1+N_c\mathcal{A}_2+N_c\mathcal{A}_3+N_c^2\mathcal{A}_4+N_c^3\mathcal{A}_5+N_c^2\mathcal{A}_6=N_c^2\langle Q(\x_3,\x'_3,\x_1,\x'_1)D(\x_2,\x'_2)\rangle,\nonumber\\
&&N_c\mathcal{A}_1+N_c^2\mathcal{A}_2+N_c^2\mathcal{A}_3+N_c\mathcal{A}_4+N_c^2\mathcal{A}_5+N_c^3\mathcal{A}_6=N_c\langle X(\x_1,\x'_1,\x_2,\x'_2,\x_3,\x'_3)\rangle,\nonumber\\
&&N_c\mathcal{A}_1+N_c^2\mathcal{A}_2+N_c^2\mathcal{A}_3+N_c^3\mathcal{A}_4+N_c^2\mathcal{A}_5+N_c\mathcal{A}_6=N_c\langle X(\x_1,\x'_1,\x_3,\x'_3,\x_2,\x'_2)\rangle,
\end{eqnarray}
with
\begin{eqnarray}
D(\x_1,\x'_1)&\equiv&\frac{1}{N_c}Tr\Big[U^\dagger(\x_1)U(\x'_1)\Big],\nonumber\\
Q(\x_1,\x'_1,\x_2,\x'_2)&\equiv&\frac{1}{N_c}Tr \Big[U^\dagger(\x_1)U(\x'_1)U^\dagger(\x_2)U(\x'_2)\Big], \nonumber\\
X(\x_1,\x'_1,\x_2,\x'_2, \x_3,\x'_3) &\equiv&\frac{1}{N_c}Tr \Big[U^\dagger(\x_1)U(\x'_1)U^\dagger(\x_2)U(\x'_2) U^\dagger(\x_3)U(\x'_3)\Big],
\end{eqnarray}
where $D$, $Q$ and $X$ are the traces of two (dipole),  four (quadrupole) and six (sextupole) light-like fundamental Wilson lines
in the background of the color fields of the target respectively.

The solution is:
\begin{eqnarray}\label{an}
\mathcal{A}_1&=&\frac{1}{(N_c^2-1)(N_c^2-4)} \Bigg[
N_c^4 \langle D(\x_1,\x'_1)D(\x_2,\x'_2)D(\x_3,\x'_3)\rangle\nonumber\\
&-&N_c^2\Big[ 2\langle D(\x_1,\x'_1)D(\x_2,\x'_2)D(\x_3,\x'_3)\rangle+\langle D(\x_1,\x'_1) Q(\x_3,\x'_3,\x_2,\x'_2) \rangle+\langle D(\x_3,\x'_3)Q(\x_1,\x'_1,\x_2,\x'_2) \rangle\nonumber\\
&&\ \ \ \ \ +\langle D(\x_2,\x'_2)Q(\x_1,\x'_1,\x_3,\x'_3)\rangle\Big]\nonumber\\
& +&2\Big[\langle X(\x_1,\x'_1,\x_2,\x'_2, \x_3,\x'_3)\rangle+\langle X(\x_1,\x'_1, \x_3,\x'_3,\x_2,\x'_2)\rangle\Big]\Bigg],\nonumber\\
 \mathcal{A}_2&=&-\frac{N_c}{(N_c^2-1)(N_c^2-4)} \Bigg[ N_c^2
 \Big[ \langle D(\x_1,\x'_1)D(\x_2,\x'_2)D(\x_3,\x'_3)\rangle-\langle D(\x_1,\x'_1)Q(\x_2,\x'_2,\x_3,\x'_3)\rangle\Big]\nonumber\\
  & +&2\langle D(\x_1,\x'_1)Q(\x_3,\x'_3,\x_2,\x'_2)\rangle - 2\langle D(\x_3,\x'_3)Q(\x_1,\x'_1,\x_2,\x'_2) \rangle
 -2 \langle D(\x_2,\x'_2)Q(\x_1,\x'_1,\x_3,\x'_3)\rangle\nonumber\\
 &+&\langle X(\x_1,\x'_1,\x_3,\x'_3, \x_2,\x'_2)\rangle+\langle X(\x_1,\x'_1,\x_2,\x'_2, \x_3,\x'_3 )\rangle\Big],\nonumber\\
 \mathcal{A}_3&=&-\frac{N_c}{(N_c^2-1)(N_c^2-4)} \Big[ N_c^2\big[\langle D(\x_1,\x'_1)
   D(\x_2,\x'_2)D(\x_3,\x'_3)\rangle -\langle D(\x_3,\x'_3)Q(\x_1,\x'_1,\x_2,\x'_2) \rangle\big]\nonumber\\
 &  +&2 \langle D(\x_3,\x'_3)Q(\x_1,\x'_1,\x_2,\x'_2)\rangle -2\langle D(\x_1,\x'_1)Q(\x_3,\x'_3,\x_2,\x'_2) \rangle
 -2\langle D(\x_2,\x'_2)Q(\x_1,\x'_1,\x_3,\x'_3)\rangle\nonumber\\
 &+&\langle X(\x_1,\x'_1,\x_3,\x'_3, \x_2,\x'_2)\rangle+\langle X(\x_1,\x'_1,\x_2,\x'_2, \x_3,\x'_3)\rangle \Big],\nonumber\\
 \mathcal{A}_4&=&
  \frac{1}{(N_c^2-1)(N_c^2-4)} \Bigg[
N_c^2\Big[ 2\langle D(\x_1,\x'_1)D(\x_2,\x'_2)D(\x_3,\x'_3)\rangle  -\langle D(\x_1,\x'_1) Q(\x_3,\x'_3,\x_2,\x'_2)\rangle \nonumber\\
&-& \langle D(\x_3,\x'_3)Q(\x_1,\x'_1,\x_2,\x'_2)\rangle 
 -\langle D(\x_2,\x'_2)Q(\x_1,\x'_1,\x_3,\x'_3)\rangle+\langle X(\x_1,\x'_1,\x_3,\x'_3, \x_2,\x'_2)\rangle \Big]\nonumber\\
 &-&2\Big[\langle X(\x_1,\x'_1,\x_3,\x'_3, \x_2,\x'_2)\rangle-\langle X(\x_1,\x'_1, \x_2,\x'_2,\x_3,\x'_3)\rangle\Big]  \Bigg],\nonumber\\
  \mathcal{A}_5&=&-\frac{N_c}{(N_c^2-1)(N_c^2-4)} \Big[
N_c^2 \big[\langle D(\x_1,\x'_1)D(\x_2,\x'_2)D(\x_3,\x'_3)\rangle-\langle D(\x_2,\x'_2)Q(\x_1,\x'_1,\x_3,\x'_3)\rangle\big]\nonumber\\
 &+&2\langle D(\x_2,\x'_2)Q(\x_1,\x'_1,\x_3,\x'_3)\rangle-2\langle D(\x_1,\x'_1) Q(\x_3,\x'_3,\x_2,\x'_2)\rangle -2\langle D(\x_3,\x'_3)Q(\x_1,\x'_1,\x_2,\x'_2) \rangle
 \nonumber\\
 &+&\langle X(\x_1,\x'_1,\x_2,\x'_2, \x_3,\x'_3)\rangle+\langle X(\x_1,\x'_1,\x_3,\x'_3, \x_2,\x'_2)\rangle\Big],\nonumber\\
 \mathcal{A}_6&=& \frac{1}{(N_c^2-1)(N_c^2-4)} \Bigg[
N_c^2\Big[ 2\langle D(\x_1,\x'_1)D(\x_2,\x'_2)D(\x_3,\x'_3)\rangle  -\langle D(\x_1,\x'_1) Q(\x_3,\x'_3,\x_2,\x'_2)\rangle \nonumber\\
&-& \langle D(\x_3,\x'_3)Q(\x_1,\x'_1,\x_2,\x'_2) \rangle
 -\langle D(\x_2,\x'_2)Q(\x_1,\x'_1,\x_3,\x'_3)\rangle+\langle X(\x_1,\x'_1,\x_2,\x'_2, \x_3,\x'_3)\rangle \Big]\nonumber\\
 &-&2\Big[\langle X(\x_1,\x'_1,\x_2,\x'_2, \x_3,\x'_3)\rangle-\langle X(\x_1,\x'_1, \x_3,\x'_3,\x_2,\x'_2)\rangle\Big]\Bigg],\
\end{eqnarray}

 Another bit of algebra gives
 \begin{eqnarray}\label{3i}
\mathcal{I}&\propto&\int_{\p_i,\p'_i,\x_i}e^{i[(\p'_1-\q_1)\x_1+(\q_1-\p_1)\x'_1+(\p'_2-\q_2)\x_2+(\q_2-\p_2)\x'_2] + (\p'_3-\q_3)\x_3+(\q_3-\p_3)\x'_3] }\nonumber\\
&\Big[&\mathcal{A}_1\sum_X A(p_1,c_1;p_2,c_2;p_3,c_3;X)A^*(p'_1,c_1;p'_2,c_2;p'_3,c_3, X),\nonumber\\
&+&\mathcal{A}_2\sum_X A(p_1,c_1;p_2,c_2;p_3,c_3;X)A^*(p'_1,c_1;p'_2,c_3;p'_3,c_2, X),\nonumber\\
&+&\mathcal{A}_3\sum_X A(p_1,c_1;p_2,c_2;p_3,c_3;X)A^*(p'_1,c_2;p'_2,c_1;p'_3,c_3, X),\nonumber\\
&+&\mathcal{A}_4\sum_X A(p_1,c_1;p_2,c_2;p_3,c_3;X)A^*(p'_1,c_2;p'_2,c_3;p'_3,c_1, X),\nonumber\\
&+&\mathcal{A}_5\sum_X A(p_1,c_1;p_2,c_2;p_3,c_3;X)A^*(p'_1,c_3;p'_2,c_2;p'_3,c_1, X),\nonumber\\
&+&\mathcal{A}_6\sum_X A(p_1,c_1;p_2,c_2;p_3,c_3;X)A^*(p'_1,c_3;p'_2,c_1;p'_3,c_2, X)\Big],\nonumber\\
&=&\int_{\p_i,\p'_i,\x_i}e^{i[(\p_1-\q_1)\x_1+(\q_1-\p'_1)\x'_1+(\p_2-\q_2)\x_2+(\q_2-\p'_2)\x'_2] + (\p_3-\q_3)\x_3+(\q_3-\p'_3)\x'_3] }\nonumber\\
&\Big[&\mathcal{A}_1T^3_{123}-\mathcal{A}_2T^3_{132}-\mathcal{A}_3T^3_{213}+\mathcal{A}_4T^3_{231}-\mathcal{A}_5T^3_{321}+
\mathcal{A}_6T^3_{312}\Big].\
\end{eqnarray}
Here we have dropped an irrelevant overall kinematic factor and have renamed the primed and unprimed momenta in the last equality for future convenience.
We  have also defined the three quark generalized parton distribution (3GTMD) as 
\begin{equation}\label{3gtmd}
T^3_{ijk}\equiv  \langle P|\psi^{\dagger a}(\p_1)\psi^a(\p'_i)\psi^{\dagger b}(\p_2)\psi^b(\p'_j) \psi^{\dagger c}(\p_3)\psi^c(\p'_k)|P\rangle,
\end{equation}
Here the spin indexes are suppressed for simplicity of notation. We will restore the various factors of $1/2$ which arise due to spin averaging in the final expressions.

The fist term in Eq.\,(\ref{3i}) clearly contains the leading in $N_c$ term, which describes independent production of the three particles. The rest of the terms  encode correlated production. 
The above expression together with Eq.\,(\ref{an}) is as far as we can go without further simplifying assumptions about the projectile 3GTMD's and the multipole scattering amplitudes. As in \cite{2pc} we will make some headway by using the properties of the scattering amplitudes as well as a simplifying approximation for 3GTMD's.

\subsection{The multipole amplitudes}
As we have discussed in \cite{2pc}, the multipole scattering amplitudes have certain properties which allow us to separate different contributions to the production cross section in terms of terms suppressed and unsuppressed by a power of the area of the projectile. 
The calculation of the production cross section involves integration over the transverse coordinates of the produced particles, which are also coordinates of the multipole amplitudes. The largest contribution therefore comes from the region of the multiple integration space where as many legs of the multipole amplitudes are as far away from each other as possible. On the other hand, we have to remember that the target ensemble is color invariant,  with the color invariance   dynamically imposed on the transverse distance scales of order of the inverse saturation momentum of the target.  In such an ensemble a multipole amplitude which has one coordinate very far way from all the others, vanishes due to color averaging.  Thus in order for the scattering amplitudes not to vanish, the points have to be at least pairwise close to each other, so that the two legs in the pair are in a color singlet. The leading (area unsuppressed) term therefore comes from the configurations where the points are pairwise close to each other, but the pairs are far away from each other in the transverse plane.
For the quadrupole amplitude it means that it is convenient to decompose it in the following way:
\begin{equation}\label{qbar}
\langle Q(1,1',2,2')\rangle=\bar Q(1,1',2,2')+\langle D(1,1')\rangle\langle D(2,2')\rangle+\langle D(1,2')\rangle\langle D(2,1')\rangle.
\end{equation}
This decomposition is such that $\bar Q(1,1',2,2')$ is only large when all points are close to each other, and thus should contribute to any cross section a term suppressed by a factor of the projectile  area. This is not to say that this term is unimportant, but it certainly has different physics associated with it. 
Similar considerations apply to the sextupole. The convenient  decomposition is
\begin{eqnarray}\label{xbar}
X(1,1',2,2',3,3')&=&\bar X(1,1',2,2',3,3')\nonumber\\
&+&D(1,1')D(2,2')D(3,3')+D(1,1')D(3,3')D(3,2')+D(1,3')D(2,2')D(3,2')\nonumber\\
&+&D(1,3')D(2,2')D(3,1')+D(1,2')D(2,1')D(3,3')\nonumber\\
&+&D(1,1')\bar Q(2,2',3,3')+D(1,3')\bar Q(2,2',3,1')+D(2,1')\bar Q(1,2',3,3')\nonumber\\
&+&D(2,2')\bar Q(1,1',3,3')
+D(3,2')\bar Q(1,1',2,3')+D(3,3')\bar Q(1,1',2,2').\
\end{eqnarray}
For simplicity of notation in Eq.\,(\ref{xbar}) and in the rest of the paper we do not indicate the averaging over the target ensemble any longer, so that from now on one should understand $D(x,y)$, $Q(x,y,u,v)$ and so on  as already averaged over the target ensemble\footnote{We note that Eq.\,(\ref{an}) contains also terms of the form $\langle D(\x_1,\x'_1) Q(\x_3,\x'_3,\x_2,\x'_2)\rangle $. These terms in general do not reduce to the product of the averages, but the correction is small at large $N_c$, e.g.  $\langle D(\x_1,\x'_1) Q(\x_3,\x'_3,\x_2,\x'_2)\rangle =\langle D(\x_1,\x'_1) \rangle\langle Q(\x_3,\x'_3,\x_2,\x'_2)\rangle+O(1/N_c^2) $, see e.g. the discussion in \cite{2pc}. Since such terms in Eq.\, (\ref{an}) are multiplied by negative powers of $N_c$, and we are only keeping terms of order $1/N_c^2$ in our final result, we will neglect these non factorizable contributions. The same comment applies to the averages of the type $\langle D(1,1')D(2,2')D(3,3')\rangle$ and the like.  It should be remembered however that in order to calculate compete area non suppressed terms at order $1/N_c^3$ and higher, these contributions have to be included. }.

This decomposition  reproduces all the leading terms that arise when all possible pairs of points that can produce  local color singlets in the amplitude are far away from each other. This refers to pairs in which one point is a quark and one is an antiquark corresponding to factors $SS^\dagger$. Whenever the adjacent points produce a factor of the type $SS$ which carries a non vanishing triality, we neglect this contribution, since such averages have extra suppression in a color invariant ensemble.
This decomposition has three types of terms. The pure dipole terms are the leading ones in terms of area dependence. The terms involving the reduced quadrupole, i.e. $\bar Q$ are suppressed by a single power of area, while we expect the term $\bar X$ to be suppressed by two powers of the area of the projectile.

\subsection{How to deal with GTMD's}
The 3GTMD's defined in Eq.\, (\ref{3gtmd}) are slightly different objects than the ones we dealt with in \cite{2pc}. Consider for example the simplest ``diagonal'' 3GTMD
\begin{equation}
T^3_D\equiv \langle P|\psi^{\dagger a}(\p_1)\psi^a(\p'_1)\psi^{\dagger b}(\p_2)\psi^b(\p'_2) \psi^{\dagger c}(\p_3)\psi^c(\p'_3)|P\rangle, 
\end{equation}
where as before $|P\rangle$ is the proton state with vanishing transverse momentum. For now we consider all quarks to be identical.
We will use the same nucleon intermediate state dominance approximation as in \cite{2pc}. The  leading $N_c$ piece in this approximation is 
\begin{equation}\label{t33}
T^3_D=\langle P|\psi^{\dagger a}(\p_1)\psi^a(\p'_1)|P,\p'_1-\p_1\rangle\langle P, \p'_1-\p_1|
\psi^{\dagger b}(\p_2)\psi^b(\p'_2)|P, \p'_1+\p'_2-\p_1-\p_2\rangle\langle |P,\p'_1+\p'_2-\p_1-\p_2|\psi^{\dagger c}(\p_3)\psi^c(\p'_3)|P\rangle.
\end{equation}
Recall that due to momentum conservation (translational invariance of the target) $\p'_3-\p_3=\p_1+\p_2-\p'_1-\p'_2$.
The middle factor in Eq.\,(\ref{t33}) is not exactly a single particle GTMD, since the latter is a matrix element between two proton state where one of the protons has zero transverse momentum. We can of course boost the whole expression so that the transverse momentum of one of the protons vanishes. This boost leads to shift in the momenta of the quark operators and not just the momenta of the proton wave functions.  However in the present paper instead we will use the approximation:
\begin{equation}
\langle P, \p'_1-\p_1|
\psi^{\dagger b}(\p_2)\psi^b(\p'_2)|P, \p'_1+\p'_2-\p_1-\p_2\rangle\approx \langle P|
\psi^{\dagger b}(\p_2)\psi^b(\p'_2)|P, \p'_2-
\p_2\rangle.
\end{equation} 
The logic here is the following. Throughout the calculation we have assumed that the momenta in the quark operators are relatively small, say of the order of several hundred Mev. This is also the typical order of magnitude of $\p'_1-\p_1$. Since the longitudinal momentum of the proton is assumed to be much larger, the transverse boost parameter needed to eliminate this transverse momentum of the proton state is very small, so that he transverse velocity involved in the boost is essentially non relativistic, ${\bf v}=1/M[\p_1-\p'_1]$ where $M$ is the proton mass. Under the same transformation the quark momentum changes by $m{\bf v}\ll \p'_1$, since the current quark mass is small $m\ll M$. Thus the change in the quark momentum is very small and we will neglect it in the following.

In the single nucleon dominance approximation we have
\begin{equation}\label{3ap}
T^3=T(\p_1;\p'_1)T(\p_2,\p'_2)T^*(\p_3; \p'_3), 
\end{equation}
where $T(\p_1,\p'_1)=\langle P|\psi^{\dagger a}(\p_1)\psi^a(\p'_1)|P,\p'_1-\p_1\rangle$ etc.  is a single quark GTMD. We have chosen to label the TMD by the momenta of the two quarks, rather than by one momentum and a momentum transfer as in \cite{2pc}.  In this equation we have suppressed  the longitudinal momentum label of the GTMD, since the focus of this paper is the transverse momentum dependence . This is not to say that the longitudinal momentum transfer is unimportant for the structure of GTMD.  It is however easily restored by simply promoting the transverse momenta to full three momenta of the operators and states involved.

We now need to understand the structure of the various 3GTMD's that appear in our master expression Eq.\,(\ref{3i}) given this single nucleon dominance approximation.The generic 3GTMD we are dealing with has the structure
\begin{equation}
T^3_{ijk}\equiv  \langle P|\psi^{\dagger a}(\p_1)\psi^a(\p'_i)\psi^{\dagger b}(\p_2)\psi^b(\p'_j) \psi^{\dagger c}(\p_3)\psi^c(\p'_k)|P\rangle.
\end{equation}
To leading order in $1/N_c$ for identical quarks we obviously have
\begin{equation}\label{leading3}
T^3_{ijk}=T(\p_1,\p'_i)T(\p_2,\p'_j)T(\p_3,\p'_k).
\end{equation}
For production of nonidentical quarks the expression Eq.\,(\ref{3ap}) has to be modified along the lines of \cite{2pc}. In particular the leading contribution is given by the product of possibly non diagonal matrix elements between distinct nucleon states:
\begin{equation}\label{leading3n}
T^3_{ijk}=T^{P\alpha}(\p_1,\p'_i)T^{\alpha\beta}(\p_2,\p'_j)T^{\beta P}(\p_3,\p'_k),
\end{equation}
with
\begin{equation}\label{ab}
T^{\alpha\beta}= \langle P_\alpha|\psi_i^{\dagger a}(\p_i)\psi_j^a(\p'_j)|P_\beta\rangle,
\end{equation}
where $|P_\alpha\rangle$ and $|P_\beta\rangle$ are the nucleon states for which the particular matrix element in Eq.\,(\ref{ab}) does not vanish given the flavors of the quarks $i$ and $j$.

 For identical particles on the other hand we need to keep sub leading in $1/N_c$ terms which are necessary to ensure the (anti)symmetry of the amplitudes. These terms lead to the Pauli blocking contribution discussed in \cite{2pc} which is also of interest to us in the present paper. To that end we should remember that the original average which gives rise to this color contraction was antisymmetric with respect to interchanges of any two momenta plus color indexes. We started with
\begin{equation}
\langle P|\psi^{\dagger {a_1}}(\p_1)\psi^{a_2}(\p'_i)\psi^{\dagger b_1}(\p_2)\psi^{b_2}(\p'_j) \psi^{\dagger c_1}(\p_3)\psi^{c_2}(\p'_k)|P\rangle, 
\end{equation}
which is obviously antisymmetric under permutations. The leading $N_c$ expression Eq.\,(\ref{leading3}) is equivalent to assuming
\begin{eqnarray}
&&\langle P|\psi^{\dagger {a_1}}(\p_1)\psi^{a_2}(\p'_i)\psi^{\dagger b_1}(\p_2)\psi^{b_2}(\p'_j) \psi^{\dagger c_1}(\p_3)\psi^{c_2}(\p'_k)|P\rangle\\
&&\approx
\langle P|\psi^{\dagger {a_1}}(\p_1)\psi^{a_2}(\p'_i)|P\rangle\langle P|\psi^{\dagger b_1}(\p_2)\psi^{b_2}(\p'_j)|P\rangle\langle P| \psi^{\dagger c_1}(\p_3)\psi^{c_2}(\p'_k)|P\rangle\nonumber\\
&&\approx\frac{1}{N_c^3}T(\p_1,\p'_i)T(\p_2,\p'_j)T(\p_3,\p'_k)\delta_{a_1a_2}\delta_{b_1b_2}\delta_{c_1c_2}.
\end{eqnarray}
To make this consistent with the (anti)commutativity of the annihilation operators, we generalize it in the following natural way
\begin{eqnarray}
&&\langle P|\psi^{\dagger {a_1}}(\p_1)\psi^{a_2}(\p'_i)\psi^{\dagger b_1}(\p_2)\psi^{b_2}(\p'_j) \psi^{\dagger c_1}(\p_3)\psi^{c_2}(\p'_k)|P\rangle\nonumber\\
&&=\frac{1}{N_c^3}\Bigg[T(\p_1,\p'_i)T(\p_2,\p'_j)T(\p_3,\p'_k)\delta_{a_1a_2}\delta_{b_1b_2}\delta_{c_1c_2}-T(\p_1,\p'_i)T(\p_2,\p'_k)T(\p_3,\p'_j)\delta_{a_1a_2}\delta_{b_1c_2}\delta_{c_1b_2}\nonumber\\
&&-T(\p_1,\p'_j)T(\p_2,\p'_i)T(\p_3,\p'_k)\delta_{a_1b_2}\delta_{b_1a_2}\delta_{c_1c_2}
+T(\p_1,\p'_j)T(\p_2,\p'_k)T(\p_3,\p'_i)\delta_{a_1b_2}\delta_{b_1c_2}\delta_{c_1a_2}\nonumber\\
&&-T(\p_1,\p'_k)T(\p_2,\p'_j)T(\p_3,\p'_i)\delta_{a_1c_2}\delta_{b_1b_2}\delta_{c_1a_2}+T(\p_1,\p'_k)T(\p_2,\p'_i)T(\p_3,\p'_j)\delta_{a_1c_2}\delta_{b_1a_2}\delta_{c_1b_2}\Bigg].\
\end{eqnarray}
Taking the appropriate trace we have
\begin{eqnarray}\label{3id}
&&T^3_{ijk}=T(\p_1,\p'_i)T(\p_2,\p'_j)T(\p_3,\p'_k)\nonumber\\
&&-\frac{1}{N_c}\Big[T(\p_1,\p'_i)T(\p_2,\p'_k)T(\p_3,\p'_j)+T(\p_1,\p'_j)T(\p_2,\p'_i)T(\p_3,\p'_k)+T(\p_1,\p'_k)T(\p_2,\p'_j)T(\p_3,\p'_i)\Big]\nonumber\\
&&+\frac{1}{N_c^2}\Big[T(\p_1,\p'_j)T(\p_2,\p'_k)T(\p_3,\p'_i)+T(\p_1,\p'_k)T(\p_2,\p'_i)T(\p_3,\p'_j)\Big].\
\end{eqnarray}
We will use this approximation to the 3GTMD in the rest of this section, and its natural generalization for nGTMD later in the paper.

Clearly the most generic case is when some of the produced quarks are identical and some are not. This can be analyzed along similar lines by combining Eqs.\,(\ref{leading3n}) and  (\ref{3id}) for appropriate quark flavors. In the rest of this paper we will not dwell on this case which is combinatorially more complicated. In fact in the following we will only consider the case when all the particles are identical. With a little extra work inferring the expression for the generic case from our results should be straightforward. 

Finally we have to restore the effects of spin in our expressions. In the approximation of the dominance of the intermediate nucleon state the basic objects we have to deal with are polarization dependent GTMD's which depend on polarization of the quarks as well as the polarization of the proton states
\begin{equation}
T^{\lambda\bar\lambda}_{s\bar s}(\p_1,\p'_i)\equiv\langle P,s|\psi_\lambda^{\dagger {a}}(\p_1)\psi_{\bar \lambda}^{a}(\p'_i)|P,\bar s\rangle.
\end{equation}
This object can in principle be decomposed into irreducible representations with respect to the little group of the fast moving quark, which leads to the appearance of several polarized GTMD's. This would complicate our expressions considerably. Since our main interest here is the effect of quantum interference and not the effects of spin, we choose to use the same simplifying assumption as in \cite{2pc}. Namely we will assume that the single particle spin averages over the proton state do not depend on the proton polarization and are dominated by the spin singlet average of the quark, i.e.
\begin{equation}
T^{\lambda\bar\lambda}_{s\bar s}(\p_1,\p'_i)=
\delta_{\lambda\bar \lambda}T(\p_1,\p'_i).
\end{equation}
We do not have a physical argument why this should be a good approximation and will content ourselves with the consideration of simplicity.

 To leading order in energy, the high energy scattering does not affect the spin of the propagating particle. Thus the polarization of a quark with momentum $\p'_i$ in the conjugate amplitude is always the same as that of the same quark (with momentum $\p_i$) in the amplitude. The cross section is traced over the polarizations  of all quarks. The polarization indexes entering any GTMD can be identified following the momentum label of the two quarks. It therefore follows that every time we have an interchange of two quarks in our expression we lose a factor of $2$  since we loose one trace over polarizations. Thus we should substitute in our expressions
 \begin{eqnarray}
&& T(\p_1,\p'_2)T(\p_2,\p'_1)\rightarrow \frac{1}{2} T(\p_1,\p'_2)T(\p_2,\p'_1)\nonumber\\
&& T(\p_1,\p'_3)T(\p_2,\p'_1)T(\p_3,\p'_2)\rightarrow \frac{1}{4}T(\p_1,\p'_3)T(\p_2,\p'_1)T(\p_3,\p'_2)\ ; \ \ \ \ etc.\
\end{eqnarray}

\subsection{Fourier transforms and notations}
Since the expression for the cross section involves Fourier transforms, we define momenta
\begin{equation}
\bar\p_i\equiv \p_i-\q_i;\ \ \ \ \ \bar\p'_i=\p'_i-\q_i. 
\end{equation}
Since the coordinates $\x_1,\ \x_2,\ \x_3$ appear in the conjugate amplitude while $\x'_1,\ \x'_2,\ \x'_3$ in the amplitude, the 
momentum conjugate to $\x_1$ is $\bar \p_1$, to $\x'_1$ is $-\bar\p'_1$ and so on (recall our renaming $\p_i\leftrightarrow \p'_i$ in Eq.\,(\ref{3i})).

To save a little bit of space we introduce concise notations for the projectile TMD's
\begin{equation}
\{i,j'\}\equiv T(\p_i,\p'_j).\end{equation}
As  for the multipole scattering amplitudes, we will denote those similarly by
\begin{equation}
[1,1']\equiv D(\x_1,\x'_1);\ \ \ \ [1,1',2,2']\equiv \bar Q(\x_1,\x'_1.\x_2,\x'_2); \ \ \ \ \ \ etc.
\end{equation}
With a slight abuse of notations we will use the same symbols to denote the Fourier transform of the amplitude with coordinates being the momenta $\bar \p_i$ and $\bar \p'_i$ defined above, i.e.
\begin{equation}
[1,1']=D(\bar\p_1,-\bar\p'_1),\ \ \ \ etc.
\end{equation}

\subsection{Expanding in $1/N_c$}
With these preliminaries we now write down Eq.\,(\ref{3i}) in an explicit form in terms of TMD's. We organize the terms in expansion in powers of $1/N_c$ except for the overall normalization factor, which we keep exact. We only write here the terms up to $O(1/N_c^2)$
since at this order we first encounter the terms contributed by the sextupole $X$ which are not present for the two quark case analyzed in \cite{2pc}.

Thus
\begin{eqnarray}
&&I_0=\frac{N_c^4}{(N_c^2-1)(N_c^2-4)}[1,1'][2,2'][3,3']\{1,1'\}\{2,2'\}\{3,3'\},\nonumber\\
&&I_1=-\frac{N_c^3}{2(N_c^2-1)(N_c^2-4)}\Bigg[ \Big[[1,1'][2,2'][3,3']+[1,1'][2,3'][2,1']+[1,1'][2,2',3,3']\Big]\{1,1'\}\{2,3'\}\{3,2'\}\nonumber\\
&&\ \ \ \ \ \ \ \ \ \ \ \  +\Big[[1,1'][2,2'][3,3']+[1,2'][2,1'][3,3']+[3,3'][1,1',2,2'] \Big]\{1,2'\}\{2,1'\}\{3,3'\}\nonumber\\
&&\ \ \ \ \ \ \ \ \ \ \ \ +   \Big[[1,1'][2,2'][3,3']+[2,2'][1,3'][3,1']+[2,2'][1,1',3,3']\Big] \{1,3'\}\{2,2'\}\{3,1'\}\Bigg],\nonumber\\
&&I_2=\frac{N_c^2}{(N_c^2-1)(N_c^2-4)}\Bigg[ \Bigg( - 5[1,1'][2,2'][3,3']+   [1,1'][2,2',3,3']+ [2,2'][1,1',3,3']+[3,3'][1,1',2,2']\Bigg) \nonumber\\
&&\times \{1,1'\}\{2,2'\}\{3,3'\}+\frac{1}{4}\Bigg([1,1'][2,2'][3,3']+[1,1'][2,3'][3,2']+[1,2'][2,1'][3,3']+[1,3'][2,2'][3,1']\nonumber\\
&&+[1,3'][2,1'][3,2']+[1,3'][3,2',2,1']+[2,1'][1,2',3,3']+[3,2'][1,1',2,3']+[1,1'][2,2',3,3']+[2,2'][1,1',3,3']\nonumber\\
&&+[3,3'][1,1',2,2']+[1,1',2,2',3,3'] \Bigg)\{1,3'\}\{2,1'\}\{3,2'\}\nonumber\\
&&+\frac{1}{4}\Bigg([1,1'][2,2'][3,3']+[1,1'][2,3'][3,2']+[1,2'][2,3][3,3']+[1,3'][2,2'][3,1']+ [1,2'][3,1'][2,3']\nonumber\\
&& +[1,2'][2,1',3,3']+[3,1'][1,3',2,2']+[2,3'][1,1',3,2']+[1,1'][2,2',3,3']+[2,2'][1,1',3,3']+[3,3'][1,1',2,2']\nonumber\\
&&+[1,1',3,3',2,2'] \Bigg)\{1,2'\}\{2,3'\}\{3,1'\}\Bigg],\
\end{eqnarray}
where the factors $1/2$ in $I_1$ and $1/4$ in two of the terms in $I_2$ are due to polarization averaging, as discussed above.

The most interesting terms in this expression are the ones that are not suppressed by the area, namely those that involve only product of dipoles.
 We assume that the average of a product of two or more dipoles factorizes into the product of averages in accordance with our earlier discussion. Assuming translational invariance of the target we have

In general
\begin{equation}
[1,i'][2,j'][3,k']=D(\bar\p_1)D(\bar \p_2)D(\bar\p_3)\delta^2(\bar\p_1-\bar\p'_i)\delta^2(\bar\p_2-\bar\p'_j)\delta^2(\bar\p_3-\bar \p'_k).
\end{equation}
Realizing the momentum delta functions we can therefore write the terms that involve only dipole contributions as (we now expand also the normalization factor in $1/N_c$)
\begin{eqnarray}\label{3qf}
&&I_0=D(\q_1-\p_1)D(\q_2-\p_2)D(\q_3-\p_3)T(\p_1,\p_1)T(\p_2,\p_2)T(\p_3,\p_3),\nonumber\\
&&I_1=-\frac{1}{2N_c}D(\q_1-\p_1)D(\q_2-\p_2)D(\q_3-\p_3)\times\nonumber\\
&&\Bigg[T(\p_1,\p_2)T(\p_2,\p_1)T(\p_3,\p_3)+T(\p_1,\p_1)T(\p_2,\p_3)T(\p_3,\p_2)+T(\p_1,\p_3)T(\p_2,\p_2)T(\p_3,\p_1)\nonumber\\
&&+T(\p_1,\p_1)T(\p_2,\p_2+\q_3-\q_2)T(\p_3,\p_3+\q_2-\q_3)+T(\p_1,\p_1+\q_3-\q_1)T(\p_2,\p_2)T(\p_3,\p_3+\q_1-\q_3)\nonumber\\
&&+T(\p_1,\p_1+\q_2-\q_1)T(\p_2,\p_2+\q_1-\q_2)T(\p_3,\p_3)\Bigg],\nonumber\\
&&I_2=\frac{1}{4N^2_c}D(\q_1-\p_1)D(\q_2-\p_2)D(\q_3-\p_3)\times\nonumber\\
&&\Bigg[T(\p_1,\p_2)T(\p_2,\p_3)T(\p_3,\p_1)+T(\p_1,\p_3)T(\p_2,\p_1)T(\p_3,\p_2)\nonumber\\
&&+T(\p_1,\p_3+\q_2-\q_3)T(\p_2,\p_2+\q_3-\q_2)T(\p_3,\p_1)+T(\p_1,\p_2+\q_3-\q_2)T(\p_2,\p_1)T(\p_3,\p_3+\q_2-\q_3)\nonumber\\
&&+T(\p_1,\p_1+\q_2-\q_1)T(\p_2,\p_3)T(\p_3,\p_2+\q_1-\q_2)+T(\p_1, \p_3)T(\p_2,\p_2+\q_2-\q_1)T(\p_3,\p_3)\nonumber\\
&&+T(\p_1,\p_2)T(\p_2,\p_1+\q_2-\q_1)T(\p_3,\p_3+\q_3-\q_2)+T(\p_1,\p_1+\q_3-\q_1)T(\p_2,\p_3+\q_2-\q_3)T(\p_3,\p_2)\nonumber\\
&&+T(\p_1,\p_1+\q_3-\q_1)T(\p_2,\p_2+\q_1-\q_2)T(\p_3,\p_3+\q_2-\q_3)\nonumber\\
&&+T(\p_1,\p_1+\q_2-\q_1)T(\p_2,\p_2+\q_3-\q_2)T(\p_3,\p_3+\q_1-\q_3)\Bigg].\
\end{eqnarray}
It is straightforward to write out the complete expression to all orders in $1/N_c$, but we will not do it explicitly  in this paper.

\section{Inclusive multiple quark production}
Our next goal is to extend the calculation to inclusive production of an arbitrary number of quarks. The calculation albeit straightforward, is combinatorially rather complicated. We therefore will not endeavor to derive  complete expression including all possible multipole contributions. It is however  not too difficult to derive the most interesting terms , i.e. those that are not suppressed by powers of area and which dominate the physics of quantum interference. Those are the terms where all multipoles can be approximated by products of dipoles in analogy with in Eqs.\,(\ref{qbar},\ref{xbar}). We will perform the explicit calculation for the production of four quarks, and then generalize the result in a straightforward way.

\subsection{Four quark production}
For production of four quarks our starting point is the following expression foe the cross section  (where we have omitted trivial kinematic factors) 
\begin{eqnarray}
\mathcal{I}&\propto&e^{[-i\q_1(\x_1-\x'_1)-i\q_2(\x'_1-\x'_2)-i\q_3(\x_2-\x'_3)-i\q_4(\x_4-\x'_4)]}\nonumber\\
&\times&\langle[U^\dagger(\x_1)U(\x'_1)]_{a_1b_1}[U^\dagger(\x_2)U(\x'_2)]_{a_2b_2} [U^\dagger(\x_3)U(\x'_3)]_{a_3b_3} [U^\dagger(\x_4)U(\x'_4)]_{a_4b_4}\rangle \nonumber\\
&\times&T^4(\{\x_1,a_1;\x_2,a_2;\x_3,a_3;\x_4,a_4\};\{\x'_1,b_1;\x'_2,b_2;\x'_3,b_3;\x'_4,b_4\}),
\end{eqnarray}
where
\begin{eqnarray}
&&T^4(\{\x_1,a_1;\x_2,a_2;\x_3,a_3;\x_4,a_4\};\{\x'_1,b_1;\x'_2,b_2;\x'_3,b_3;\x'_4,b_4\})\nonumber\\
&\equiv&
\langle P|\psi^\dagger(\x_1,a_1)\psi^\dagger(\x_2,a_2)\psi^\dagger(\x_3,a_3)\psi^\dagger(\x_4,a_4)
\psi(\x'_4,b_4) \psi(\x'_3,b_3)\psi(\x'_2,b_2)\psi(\x'_1,b_1)|P\rangle.
\end{eqnarray}
Previously we have used the color invariance of the target field distribution to write this expression in terms of multipoles, and the factored the multipoles into dipoles and remaining terms which are suppressed by the area. Since now we are only interested in the dipole contributions, we can directly factorize the target average using pairwise contractions
\begin{equation}
\langle U^\dagger(\x)_{\alpha\beta}U(\y)_{\gamma\delta}\rangle=\delta_{\alpha\delta}\delta_{\beta\gamma}[\x,\y].
\end{equation}
These pairwise contractions generate 24 terms:
\begin{eqnarray}\label{4p}
&&I\equiv \langle[U^\dagger(\x_1)U(\x'_1)]_{a_1b_1}[U^\dagger(\x_2)U(\x'_2)]_{a_2b_2} [U^\dagger(\x_3)U(\x'_3)]_{a_3b_3} [U^\dagger(\x_4)U(\x'_4)]_{a_4b_4}\rangle\nonumber\\
&\times& T^4(\{\x_1,a_1;\x_2,a_2;\x_3,a_3;\x_4,a_4\};\{\x'_1,b_1;\x'_2,b_2;\x'_3,b_3;\x'_4,b_4\})\nonumber\\
&=&[1,1'][2,2'][3,3'][4,4']T^4_{1234}\nonumber\\
&-&\frac{1}{N}\Bigg[[1,1'][2,2'][3,4'][4,3']T^4_{1243}+[1,2'][2,1'][3,3'][4,4']T^4_{2134}+[1,1'][2,3'][3,2'][4,4']T^4_{1324}\nonumber\\
&+&[1,1'][2,4'][3,3'][4,2']T^4_{1432}+[1,3'][2,2'][3,1'][4,4']T^4_{3214}
+[1,4'][2,2'][3,3'][4,1']T^4_{4231}\Bigg]\nonumber\\
&+&\frac{1}{N^2}\Bigg[[1,2'][2,1'][3,4'][4,3']T^4_{2143}+[1,3'][2,4'][3,1'][4,2']T^4_{3412}
+[1,4'][2,3'][3,2'][4,1']T^4_{4321}
+[1,1'][2,3'][3,4'][4,2']T^4_{1342}\nonumber\\
&+&[1,1'][2,4'][3,2'][4,3']T^4_{1423}+
[1,4'][2,2'][3,1'][4,3']T^4_{4213}+[1,3'][2,2'][3,4'][4,1']T^4_{3241}+[1,2'][2,4'][3,3'][4,1']T^4_{2431}\nonumber\\
&+&[1,4'][2,1'][3,3'][4,2']T^4_{4132}+[1,2'][2,3'][3,1'][4,4']T^4_{2314}+[1,3'][2,1'][3,2'][4,4']T^4_{3124}
\Bigg]\nonumber\\
&-&\frac{1}{N^3}\Bigg[[1,2'][2,3'][3,4'][4,1']T^4_{2341}+[1,2'][2,4'][3,1'][4,3']T^4_{2413}+[1,3'][2,1'][3,4'][4,2']T^4_{3142}\nonumber\\&+&[1,3'][2,4'][3,2'][4,1']T^4_{3421}
+[1,4'][2,1'][3,2'][4,3']T^4_{4123}+[1,4'][2,3'][3,1'][4,2']T^4_{4312}\Bigg]. 
\end{eqnarray}
The signs in this expression arise from the reordering of the fermionic operators in the definition of TMD's so that each TMD is a product of color singlets, i.e
\begin{equation}\label{4gtmd}
T^4_{ijkl}\equiv\langle P|\psi^\dagger(\x_1,a_1)\psi(\x'_i,a_1)\psi^\dagger(\x_2,a_2)\psi(\x'_j,a_2)\psi^\dagger(\x_3,a_3)\psi(\x'_k,a_3)\psi^\dagger(\x_4,a_4)
\psi(\x'_l,a_4) )|P\rangle.
\end{equation}
Clearly to leading order in $1/N_c$ we have
\begin{equation}
T^4_{ijkl}\rightarrow_{N_c\rightarrow\infty}\{1,i'\}\{2,j'\}\{3,k'\}\{4,l'\}.
\end{equation}
Adhering to this approximation we will correctly account for the HBT-like terms, but will not take into account the terms due to Fermi statistics effects in the initial wave function - the Pauli blocking.

\subsection{The Pauli Blocking contribution}
To include the Pauli blocking terms we have to antisymmetrize the proton wave function. We will use the same approximation as for three quark production. We write in coordinate space
\begin{eqnarray}\label{pb4tmd}
&&T^4_{ijkl}=\{1,i'\}\{2,j'\}\{3,k'\}\{4,l'\}\nonumber\\
&&-\frac{1}{N}\Bigg[\{1,j'\}\{2,i'\}\{3,k'\}\{4,l'\}+\{1,k'\}\{2,j'\}\{3,j'\}\{4,l'\}+\{1,l'\}\{2,j'\}\{3,k'\}\{4,i'\}\nonumber\\
&&\ \ \ \ \ \ \ \ +\{1,i'\}\{2,k'\}\{3,j'\}\{4,l'\}+\{1,i'\}\{2,l'\}\{3,k'\}\{4,j'\}+\{1,i'\}\{2,j'\}\{3,l'\}\{4,k'\}\Bigg]\nonumber\\
&&+\frac{1}{N^2}\Bigg[\{1,j'\}\{2,i'\}\{3,l'\}\{4,k'\}+\{1,k'\}\{2,l'\}\{3,i'\}\{4,j'\}+\{1,l'\}\{2,k'\}\{3,j'\}\{4,i'\}\nonumber\\
&&\ \ \ \ \ \ \ \ +\{1,i'\}\{2,k'\}\{3,l'\}\{4,j'\}+\{1,i'\}\{2,l'\}\{3,j'\}\{4,k'\}+\{1,k'\}\{2,j'\}\{3,l'\}\{4,i'\}\nonumber\\
&&\ \ \ \ \ \ \ \ +\{1,l'\}\{2,j'\}\{3,i'\}\{4,k'\}
+\{1,j'\}\{2,l'\}\{3,k'\}\{4,i'\}
+\{1,l'\}\{2,i'\}\{3,k'\}\{4,j'\}\nonumber\\
&&\ \ \ \ \ \ \ \ +\{1,j'\}\{2,k'\}\{3,i'\}\{4,l'\}+\{1,k'\}\{2,i'\}\{3,j'\}\{4,l'\}\Bigg]\nonumber\\
&&-\frac{1}{N^3}\Bigg[\{1,j'\}\{2,k'\}\{3,l'\}\{4,i'\}+\{1j'\}\{2,l'\}\{3,i'\}\{4,k'\}+\{1,k'\}\{2,i'\}\{3,l'\}\{4,j'\}\nonumber\\
&&\ \ \ \ \ \ \ +\{1,k'\}\{2,k'\}\{3,l'\}\{4,i'\}+\{1,l'\}\{2,i'\}\{3,j'\}\{4,k'\}+\{1,l'\}\{2,k'\}\{3,i'\}\{4,j'\}\Bigg].\
\end{eqnarray}

Finally, just like for three quark production to account for the spin we need to introduce a factor of $1/2^n$ for every permutation of indexes $(i',j',k',l)'$ where $n$ is the number of index permutations necessary to obtain it from the trivial permutation $(1,2,3,4)$.

We can now organize all terms in orders of $1/N$. The resulting expression is fairly long and we do not write it out out exhaustively in the text. Instead we present it in Appendix B for reference. In the same appendix we give the expression for the cross section transformed to Fourier space.  Our main goal is to extend these expressions to an arbitrary number of particles. Thankfully this can be done essentially by inspection.

\subsection{Generalizing to arbitrary number of produced particles}
 We start with writing Eq.\,(\ref{4p}) is a suggestive form. Obviously the expression is a sum over all permutations of the four coordinates $i',j',k',l'$.  let us introduce the following notations: $P_{ij}$ denotes a permutation of the four coordinates in which only the $i$'th and $j$'th coordinates are interchanged, $P_{ijk}$ - a permutation where the three coordinates are interchanged with each other without leaving any  one in its place, $P_{ij;kl}$ - permutation where the coordinates are interchanged within two pairs, etc. Accordingly $P(k)_{ij}$ etc means the $k$'th number in the appropriate permutation. We can then write directly the generalization of  Eq.\,(\ref{4p}) to n particle production:
\begin{eqnarray}
&&I=\prod_{a=1}^{n}[a,a]T^n_{12...n}\nonumber\\
&&-\frac{1}{N}\sum_{P_{ij}}\prod_{a=1}^{n}[a, P(a)_{ij}]T^n_{P(1)_{ij}...P(n)_{ij}}\nonumber\\
&&+\frac{1}{N^2}\Bigg[\sum_{P_{ijk}}\prod_{a=1}^{n}[a, P(a)_{ijk}]T^n_{P(1)_{ijk}...P(n)_{ijk}}+\sum_{P_{ij;kl}}\prod_{a=1}^{n}[a, P(a)_{ij;k}]T^n_{P(1)_{ij;kl}...P(n)_{ij;kl}}\Bigg]\nonumber\\
&&-\frac{1}{N^3}\sum_{P_{ijkl}}\prod_{a=1}^{n}[a, P(a)_{ijkl}]T^n_{P(1)_{ijkl}...P(n)_{ijkl}}\nonumber\\
&&+\dots.\
\end{eqnarray}
Here we have defined the nGTMD's $T^n$ in exact analogy with Eq.\,(\ref{4gtmd}). We now use the same parametrization of nGTMD in terms of single TMD's as in Eq.\,(\ref{pb4tmd}).
We can then immediately generalize Eq.\,(\ref{full4}) 
\begin{eqnarray}
&&I_0= \prod_{a=1}^{n}[a,a]\prod_{b=1}^n\{b,b\},\nonumber\\
&&I_1=-\frac{1}{2N}\sum_{P_{ij}}\Bigg[\prod_{a=1}^{n}[a, P(a)_{ij}] + \prod_{a=1}^{n}[a,a]\Bigg] \prod_{b=1}^n\{b, P(b)_{ij}\},\nonumber\\
&&I_2=\frac{1}{4N^2}\Bigg\{\sum_{P_{ij;kl}}\Bigg[\prod_{a=1}^{n}[a, P(a)_{ij;kl}] +\prod_{a=1}^{n}[a, P(a)_{ij}]+\Pi_{a=1}^{n}[a, P(a)_{kl}]+ \Pi_{a=1}^{n}[a,a]\Bigg] \prod_{b=1}^n\{b, P(b)_{ij;kl}\}\nonumber\\
&&\sum_{P_{ijk}}\Bigg[\prod_{a=1}^{n}[a, P(a)_{ijk}] +\prod_{a=1}^{n}[a, P(a)_{ij}]+\prod_{a=1}^{n}[a, P(a)_{jk}]+\prod_{a=1}^{n}[a, P(a)_{kl}]+ \prod_{a=1}^{n}[a,a]\Bigg] \prod_{b=1}^n\{b, P(b)_{ijk}\}\nonumber\\
&&+4 \sum_{P_{ij}}\prod_{a=1}^{n}[a, P(a)_{ij}]\prod_{b=1}^{n}\{b,b\}\Bigg\}.\
\end{eqnarray}
And finally the generalization of the expression in momentum space Eq.\,(\ref{full4p}) is
\begin{equation}
I_k=\prod_{i=1}^nD(\p_i-\q_i)J_k,
\end{equation}
with
\begin{eqnarray}\label{jn}
J_0&=&\prod_{i=1}^{n}\{\p_i,\p_i\},\nonumber\\
J_1&=&-\frac{1}{2N}\sum_{\{i,j\}} \Big[\{\p_i,\p_i+\q_j-\q_i\}\{\p_j,\p_j+\q_i-\q_j\}+\{\p_i,\p_j\}\{\p_j,\p_i\}\Big]\prod_{k\ne i,j}\{\p_k,\p_k\},\nonumber\\
J_2&=&\frac{1}{4N^2}\Bigg[\sum_{\{i,j\};\{l,m\}}\{\p_i,\p_i+\q_j-\q_i\}\{\p_j,\p_j+\q_i-\q_j\}\{\p_l,\p_l+\q_m-\q_l\}\{\p_m,\p_m+\q_l-\q_m\}\prod_{k\ne i,j,l,m}\{\p_k,\p_k\}\nonumber\\
&+&\sum_{\{i,j\};\{l,m\}}\{\p_i,\p_i+\q_j-\q_i\}\{\p_j,\p_j+\q_i-\q_j\}\{\p_l,\p_m\}\{\p_m,\p_l\}\prod_{k\ne i,j,l,m}\{\p_k,\p_k\}\nonumber\\
&+&\sum_{\{i,j\};\{l,m\}}\{\p_i,\p_j\}\{\p_j,\p_i\}\{\p_l,\p_m\}\{\p_m,\p_l\}\prod_{k\ne i,j,l,m}\{\p_k,\p_k\}\nonumber\\
&+&\frac{1}{3}\sum_{\{i,j,l\}}\{\p_i,\p_i+\q_j-\q_i\}\{\p_j,\p_j+\q_l-\q_j\}\{\p_l,\p_l+\q_i-\q_l\}\prod_{k\ne i,j,l}\{\p_k,\p_k\}\nonumber\\
&+&\frac{1}{3}\sum_{\{i,j,l\}}\{\p_i,\p_j\}\{\p_j,\p_l\}\{\p_l,\p_i\}\prod_{k\ne i,j,l}\{\p_k,\p_k\}\Big]\nonumber\\
&+&\sum_{\{i,j,l\}}\{\p_i,\p_j+\q_k-\q_j\}\{\p_j,\p_i\}\{\p_l,\p_l+\q_i-\q_l\}\prod_{k\ne i,j,l}\{\p_k,\p_k\}\nonumber\\
&+&4\sum_{\{i,j\}} \{\p_i,\p_i+\q_j-\q_i\}\{\p_j,\p_j+\q_i-\q_j\}\prod_{k\ne i,j}\{\p_k,\p_k\}\Bigg].\
\end{eqnarray}
Here $\{i,j\}$ is a pair of indexes, and $\{i,j,l\}$ is an {\bf ordered} triplet of indexes, such that for example $\{1,2,3\}$ is considered distinct from $\{2,1,3\}$ and from $\{2,3,1\}$.

The meaning of the various terms in Eq.\,(\ref{jn}) is quite clear. All terms that involve TMD's that depend on the final state momenta$\q_i$ arise from the HBT - like contributions of the final state particles. On the other hand any TMD of the form $\{\p_i,\p_j\}$  with $i\ne j$ appears due the Pauli blocking effects in the projectile wave function. It is thus clear how this result is modified if not all the quarks are identical. In this case any term that has a factor $\{\p_i,\p_j\}$ where  $i$ and $j$ refer to nonidentical quarks, should be dropped. Also any TMD of the form $ \{\p_i,\p_i+\q_j-\q_i\}$ with $i$ and $j$  corresponding to nonidentical quarks should be replaced by $T^{\alpha\beta}$ defined in Eq.\,(\ref{ab}) with $|P_\alpha\rangle$ and $|P_\beta\rangle$ allowed by flavor conservation.

\section{Discussion}
In this paper we have derived formal expressions for multi quark inclusive production, keeping only terms that are not suppressed by factors of area. Our main motivation was to explore how  the effects of quantum interference affect multi particle production beyond the two particle correlations effects present already for inclusive production of two particles. 
Although our final expressions are fairly lengthy, the main properties of the cross section can be inferred without numerical calculations. In this section we would like to discuss these qualitative features.

We first concentrate on the  three particle production Eq.\,(\ref{3qf}), since to order $1/N_c^2$ no qualitatively new features appear in multiple quark production. The leading term $I_0$ clearly describes independent production of the three quarks. The next term $I_1$ is essentially the same as in the two particle production considered in \cite{2pc}. Here one particle is produced independently, while the other two are correlated via either the HBT effect or the Pauli blocking effect in the initial wave function.
 It is easy to distinguish between the ``HBT'' and ``Pauli blocking'' terms. The former involve GTMD's with momentum transfer corresponding to momentum differences of final state particles $\q_i-\q_j$, while the latter - the incoming parton momenta $\p_i-\p_j$. As in \cite{2pc} we can use a simplified form of the GTMD in order to understand the qualitative features of the correlation. We take
 \begin{equation}
 T(\p,\k)=T(\frac{\p+\k}{2})F(\k-\p); \ \ \ \ \ \ \  F(\k)=\frac{1}{\frac{\k^2}{\Lambda^2}+1}.
 \end{equation}
 Here $T(\q)$ is the TMD and  the form factor$F$  suppresses momentum transfer larger than some soft hadronic scale $\Lambda$, which naturally has the meaning of the inverse radius of the proton. Thus the various HBT terms in $I_1$ suppress particle production when  momenta 
of any two particles are within the distance $\Lambda$ of each other. There is also suppression of production from incoming particles with similar momenta due to the Pauli blocking effect. This effect is different in the sense that as a result of the interaction with the target  the incoming momenta are smeared by the amount of  
the order of the saturation momentum of the target $Q_s$. Thus the width of the trough in correlation of the {\bf  produced} momentum is of order $Q_s$ and not $\Lambda$ as in the case of HBT.  

One should keep in mind this significant difference between the HBT induced and the Pauli bocking induced correlations\cite{bosee}. 
The HBT correlations are due to interference of the signals emitted by {\it incoherent} emitters from the surface of the proton right after scattering, see for example discussion in \cite{bosee}. 
The width of the HBT correlated signal in momentum space is determined by the inverse size of the proton and  is approximately independent of the momentum of produced particles. It is thus present at large transverse momenta $\q_i$.  
The value of the saturation momentum of the target place a secondary role as long as it is large enough. However for small $Q_s$ the magnitude of the HBT signal must be suppressed, since in this case the number of uncorrelated emitters is small.

On the other hand the Pauli blocking directly correlates momenta of partons in the proton wave function. The momenta transferred from the target to the two partons are not correlated. If $Q_s$ of the target is greater than the average transverse momentum of the incoming partons we expect the correlations present in the initial wave function to be washed away by the scattering. In this regime we thus expect the Pauli blocking terms to contribute to the isotropic ``pedestal'' but not to angularly correlated signal.
This property in fact can be explicitly seen from our formulae. Consider for example one of the Pauli blocking terms in Eq.\,(\ref{3qf})
\begin{equation}
\int_{\p_i}D(\q_1-\p_1)D(\q_2-\p_2)D(\q_3-\p_3)T(\p_1,\p_2)T(\p_2,\p_1)T(\p_3,\p_3).
\end{equation}
For large transverse momenta, which are much larger than possible intrinsic momenta in the proton wave function $\q_i\gg\p_i$, one can neglect the transverse momenta $\p_i$ in the argument of the dipole amplitudes $D(\q_i-\p_i)$. We then get
\begin{equation}
D(\q_1)D(\q_2)D(\q_3)\int_{\p_i}T(\p_1,\p_2)T(\p_2,\p_1)T(\p_3,\p_3).
\end{equation}
This clearly does not induce any angular correlations between the produced particles, although it does gives a negative isotropic contribution to the inclusive three particle production.

 On the other hand  if $Q_s$ is smaller than the intrinsic transverse momentum of the projectile, the Pauli blocking correlations should be directly observed as angular correlations in the emission.

Thus for an asymmetric collision, like p-A we expect the HBT signal to dominate at large $\q_i$ for emission in the direction of the dilute object (proton), while the Pauli blocking signal to dominate in the direction of the nucleus. It is possible that at mid rapidity both contributions are comparable. One has to keep in mind though that our present framework is quantitatively correct only for the forward production in the proton direction and thus a discussion of correlations at mid rapidity is strictly speaking outside the scope of the present paper.

The terms contained in $I_1$ contribute to correlations between two particles only, while the third particle in any of these terms is emitted independently. These terms do not contribute to three particle collectivity measures, such as $v_2^3$ since the emission angle of one of the particles is isotropic with respect to the other two. In this respect the first interesting term is $I_2$. It is thus interesting to understand what kind of three particle correlation it induces. First off we note that all the terms in $I_2$ are positive. Therefore they lead to partial compensation of the negative correlation due to ``pairwise'' HBT and Pauli blocking.  There are three types of terms in $I_2$. The first type is
\begin{equation}
\int_{\p_i}D(\q_1-\p_1)D(\q_2-\p_2)D(\q_3-\p_3)T(\p_1,\p_2)T(\p_2,\p_3)T(\p_3,\p_1).
\end{equation}
This is a kind of ``unitarization'' correction to pairwise Pauli blocking. We interpret it as an indication that $I_1$ ``over subtracts'' the Pauli blocking contribution in the regime when all three quarks in the proton wave function have equal momenta. The positive contribution from $I_2$ rectifies this ``over subtraction''.

The second type of term is proportional to
\begin{equation}
\int_{\p_i}D(\q_1-\p_1)D(\q_2-\p_2)D(\q_3-\p_3)T(\p_1,\p_1+\q_3-\q_1)T(\p_2,\p_3+\q_2-\q_3)T(\p_3,\p_2).
\end{equation}
This is a ``mixed'' HBT-Pauli blocking correction. The magnitude of this term is maximal when $\q_3=\q_1=\q_2$ and $\p_2=\p_3$. Interestingly, although this term clearly requires particles 2 and 3 to be identical, it does contribute to angular correlations even if we neglect the $\p$ dependence of the dipole amplitudes.

Finally the last type of terms is exemplified by
\begin{equation}
\int_{\p_i}D(\q_1-\p_1)D(\q_2-\p_2)D(\q_3-\p_3)T(\p_1,\p_1+\q_2-\q_1)T(\p_2,\p_2+\q_3-\q_2)T(\p_3,\p_3+\q_1-\q_3).
\end{equation}
This again has a simple interpretation as a ``unitarization correction''. This time the correction is to the pairwise HBT interference term.

For production of more than 3 quarks, the only difference is that at order $1/N^2_c$  one also has terms that  correlate two pairs of quarks, again by either HBT or PB mechanisms. In fact he pattern is quite clear and one can straightforwardly generalize Eq.\, (\ref{jn}) to higher order terms in $1/N_c$.

We note, that in the imaginary world where the quarks were scalar particles, the calculation would have been almost identical. The only difference (except for the absence of the factors of $1/2$ that appeared due to spin averaging) is that  all the interference  terms would be positive. Thus not only the interference effects would enhance rather than deplete the ``same side'' particle emission, the higher order in $1/N_c$ terms  would add to the effect rather than suppress it.

Another point worth mentioning is that although we formally expand in powers of $1/N_c$, this expansion quickly becomes unreliable for large number of produced particles $n$.  Even if we count only the ``HBT'' type terms, it is clear from Eq.\, (\ref{jn}) that $J_1$ is a sum of $\frac{n(n-1)}{2}\sim O(n^2)$ such terms, while $J_2$ is a sum of $\frac{n(n-1)(n-2)(n-3)}{4}+\frac{n(n-1)(n-2)}{3!}\sim O(n^4)$ terms. Already at $n\sim \sqrt{N_c}$ the number of terms in all $J_m$ compensates for the suppression factor $1/N^m_c$.
Thus one has to study the unexpanded expression rather than fixed order in $1/N_c$. Since the terms in the series have alternating signs it would be interesting to see what is the nett effect at large $n$. Again we note that for scalar quarks where all the terms are positive, the effect at large $n$ is positive and can be very large. 

Finally we iterate that our result for  $n=3$ is complete, while for multiple production ($n>3$) it collect only the terms which are not suppressed by powers of area. In the parlance of ref.\cite{bmu} those are terms leading in the number of sources. These terms are responsible for the quantum interference effects. The recent impressive calculation of two quark production  \cite{mace} within the MV model \cite{mv} does not include these terms.  The  starting point of \cite{mace} is the expression for the two particle production given as the target average of product of two dipoles. This target average is then carefully calculated within the MV model including non factorisable $1/N_c^2$ suppressed terms terms. This corresponds to a careful target averaging of our term $I_0$. The corrections that ref.\cite{mace} takes into account therefore is what we have called classical terms which arise from the part of the phase space when more than two points in the product of two dipoles are within the distance of $1/Q_s$. As noted in \cite{2pc} such terms are suppressed by at least the factor of $1/N_c^2$, and are therefore sub leading relative to $I_1$ and are of the same order as $I_2$. Clearly a calculation of particle correlations within the CGC approach is incomplete without including the quantum interference terms.

\appendix\section{The coefficient functions}

Using the decomposition of the multipoles in terms of the localized amplitudes we can rewrite the coefficient functions  defined in Section II. To do that first of all we introduce a concise notation:
\begin{equation}
[1,1']\equiv D(\x_1,\x'_1);\ \ \ \ [1,1',2,2']\equiv \bar Q(\x_1,\x'_1.\x_2,\x'_2); \ \ \ \ \ \ etc.
\end{equation}
With this we can write
\begin{eqnarray}
\mathcal{A}_1&=&\frac{1}{(N_c^2-1)(N_c^2-4)} \Bigg[
N_c^4[1,1'][2,2'][3,3']\\
&-&N_c^2\Big[ 5[1,1'][2,2'][3,3']+
[1,1'][2,3'][3,2']+[1,2'][2,1'][3,3']+[1,3'][2,2'][3,1']\\
&
+&[1,1'][2,2',3,3'] + [3,3'][1,1',2,2'] 
 +[2,2'][1,1',3,3']\Big]\nonumber\\
 &+&4[1,1'][2,2',][3,3'] +4[1,1'][2,3'][3,2']+4[1,3'][3,1'][2,2']+4[1,2'][2,1'][3,3']+2[1,3'][2,1'][3,2']+2[1,2'][3,1'][2,3']\nonumber\\
 &+&
 2\Big[2[1,1'][2,2',3,3']+2[2,2',][1,1',3,3']+2[3,3'][1,1',2,2']\nonumber\\
 &+&[1,3'][2,2',3,1']+[1,2'][2,1',3,3']+[2,1'][1,2',3,3']+[2,3'][1,1',3,2']+[3,1'][1,3',2,2']+2[3,2'][1,1',2,3']
 \Big]\nonumber\\
 &+&2\Big[[1,1',2,2',3,3']+[1,1',3,3',2,2'] \Big]\Bigg],\nonumber\\
 \mathcal{A}_2&=&\frac{1}{(N_c^2-1)(N_c^2-4)} \Bigg[ N_c^3
 \Big[[1,1'][2,3'][3,2']+[1,1'][2,2',3,3']\Big]\nonumber\\
 &+&N_c\Big[4[1,1'][3,2'][2,3']+[1,2'][3,1'][2,3']+[1,3'][2,1'][3,2']\nonumber\\
 &+&4[1,1'][2,2',3,3']+[1,2'][2,1',3,3']+[1,3'][2,2',3,1']+[3,1'][1,3',2,2']+[3,2'][1,1',2,3']+[2,1'][1,2',3,3']
\nonumber\\
 &+&[2,3'][1,1',3,2']+[1,1',3,3',2,2']+[1,1',2,2',3,3'] \Big]\Bigg],\nonumber\\
 \mathcal{A}_3&=&\frac{1}{(N_c^2-1)(N_c^2-4)} \Bigg[ N_c^3\Big[[1,2'][2,1'][3,3']+[3,3'][1,1',2,2'] \Big]\nonumber\\
 &  -&N_c\Big[4[1,2'][2,1'][3,3'] +[1,3'][2,1'][3,2']+[1,2'][3,1'][2,3']\nonumber\\
 &+&4[3,3'][1,1',2,2']+[1,3'][2,2',3,1']+[1,2'][2,1',3,3']+[2,1'][1,2',3,3']+[2,3'][1,1',3,2']+[3,1'][1,3',2,2']
 \nonumber\\
 &+&[3,2'][1,1',2,3']+[1,1',2,2',3,3']+[1,1',3,3',2,2']\Big] \Bigg],\nonumber\\
 \mathcal{A}_4&=&\frac{1}{(N_c^2-1)(N_c^2-4)} \Bigg[
N_c^2\Big[ [1,2'][3,1'][2,3']+[1,2'][2,1',3,3']+[3,1'][1,3',2,2']+[2,3'][1,1',3,2']+[1,1',3,3',2,2']
\Big]\nonumber\\
 &+&2\Big[[1,3'][2,1'][3,2']-[1,2'][3,1'][2,3']\nonumber\\
 & +&[1,3'][2,2',3,1']-[1,2'][2,1',3,3']+[2,1'][1,2',3,3']-[2,3'][1,1',3,2']+[3,2'][1,1',2,3']-[3,1'][1,3',2,2'] \nonumber\\
 &+&[1,1',2,2',3,3']-[1,1',3,3',2,2']  \Big]\Bigg],\nonumber\\
 \mathcal{A}_5&=&\frac{1}{(N_c^2-1)(N_c^2-4)} \Bigg[ N_c^3
 \Big[[2,2'][1,3'][3,1']+[2,2'][1,1',3,3']\Big]\nonumber\\
 &+&N_c\Big[4[2,2'][3,1'][1,3']+[2,1'][3,2'][1,3']+[2,3'][1,2'][3,1']\nonumber\\
 &+&4[2,2'][1,1',3,3']+[2,1'][1,2',3,3']+[2,3'][1,1',3,2']+[3,2'][1,1',2,3']+[3,1'][2,2',1,3']+[1,2'][3,2,3,3']
\nonumber\\
 &+&[1,3'][2,2',3,1']+[1,1',3,3',2,2']+[1,1',2,2',3,3'] \Big]\Bigg],\nonumber\\
 \mathcal{A}_6&=& \frac{1}{(N_c^2-1)(N_c^2-4)} \Bigg[
N_c^2\Big[ [1,3'][2,1'][3,2']+[1,3'][2,1',3,3']+[2,1'][1,2',3,3']+[3,2'][1,1',2,3']+[1,1',2,2',3,3']
\Big]\nonumber\\
 &-&2\Big[[1,3'][2,1'][3,2']-[1,2'][3,1'][2,3']\nonumber\\
 & +&[1,3'][2,2',3,1']-[1,2'][2,1',3,3']+[2,1'][1,2',3,3']-[2,3'][1,1',3,2']+[3,2'][1,1',2,3']-[3,1'][1,3',2,2'] \nonumber\\
 &+&[1,1',2,2',3,3']-[1,1',3,3',2,2']  \Big]\Bigg],\nonumber\\
\end{eqnarray}

\section{The four quark production}
In this appendix we collect the expressions for the cross section for production of four identical quarks.

The following refers to the expansion of Eq.\,(\ref{4p}) in powers of $1/N$ assuming all particles are identical.
\begin{eqnarray}\label{full4}
&&I_0=[1,1'][2,2'][3,3'][4,4']\{1,1'\}\{2,2'\}\{3,3'\}\{4,4'\},\nonumber\\
&&I_1=-\frac{1}{2N}\Bigg[\Big([1,1'][2,2'][2,2''][4,3']  +[1,1'][2,2'][3,3'][4,4']\Big)\{1,1'\}\{2,2'\}\{3,4'\}\{4,3'\}\nonumber\\
&&+\Big([1,2'][2,1'][3,3'][4,4']+[1,1'][2,2'][3,3'][4,4'] \Big) \{1,2'\}\{2,1'\}\{3,3'\}\{4,4'\}\nonumber\\
&&+\Big([1,1'][2,3'][3,2'][4,4']+[1,1'][2,2'][3,3'][4,4']\Big) \{1,1'\}\{2,3'\}\{3,2'\}\{4,4'\} \nonumber\\
&&+\Big([1,1'][2,4'][3,3'][4,2']+[1,1'][2,2'][3,3'][4,4']\Big)\{1,1'\}\{2,4'\}\{3,3'\}\{4,2'\}\nonumber\\
&&+\Big([1,3'][2,2'][3,1'][4,4']+[1,1'][2,2'][3,3'][4,4']\Big)\{1,3'\}\{2,2'\}\{3,1'\}\{4,4'\}\nonumber\\
&&+\Big([1,4'][2,2'][3,3'][4,1']+[1,1'][2,2'][3,3'][4,4']\Big)\{1,4'\}\{2,2'\}\{3,3'\}\{4,1'\}\Bigg],\nonumber\\
&&I_2=\frac{1}{4N^2}\Bigg[\Big([1,2'][2,1'][3,4'][4,3']+[1,1'][2,2'][3,4'][4,3']+[1,2'][2,1'][3,3'][4,4']\Big)\{1,2'\}\{2,1'\}\{3,4'\}\{4,3'\}\nonumber\\
&&+\Big([1,3'][2,4'][3,1'][4,2']+[1,1'][2,4'][3,3'][4,2']+[1,3'][2,2'][3,1'][4,4'\Big)\{1,3'\}\{2,4'\}\{3,1'\}\{4,2'\}\nonumber\\
&&+\Big([1,4'][2,3'][3,2'][4,1']+[1,1'][2,3'][3,2'][4,4']+[1,4'][2,2'][3,3'][4,1']\Big)]\{1,4'\}\{2,3'\}\{3,2'\}\{4,1'\}\nonumber\\
&&+\Big([1,1'][2,3'][3,4'][4,2']+[1,1'][2,2'][3,4'][4,3']+[1,1'][2,3'][3,2'][4,4']+[1,1'][2,4'][3,3'][4,2']\Big)\{1,1'\}\{2,3'\}\{3,4'\}\{4,2'\}\nonumber\\
&&+\Big([1,1'][2,4'][3,2'][4,3']+[1,1'][2,2'][3,4'][4,3']+[1,1'][2,3'][3,2'][4,4']+[1,1'][2,4'][3,3'][4,2']\Big)\{1,1'\}\{2,4'\}\{3,2'\}\{4,3'\}\nonumber\\
&&+
\Big([1,4'][2,2'][3,1'][4,3']+[1,1'][2,2'][3,4'][4,3']+[1,3'][2,2'][3,1'][4,4']+[1,4'][2,2'][3,3'][4,1']\Big)\{1,4'\}\{2,2'\}\{3,1'\}\{4,3'\}\nonumber\\
&&+\Big([1,3'][2,2'][3,4'][4,1']+[1,1'][2,2'][3,4'][4,3']+[1,3'][2,2'][3,1'][4,4']+[1,4'][2,2'][3,3'][4,1']\Big)\{1,3'\}\{2,2'\}\{3,4'\}\{4,1'\}\nonumber\\
&&+\Big([1,2'][2,4'][3,3'][4,1']+[1,2'][2,1'][3,3'][4,4']+[1,1'][2,4'][3,3'][4,2']+[1,4'][2,2'][3,3'][4,1']\Big)\{1,2'\}\{2,4'\}\{3,3'\}\{4,1'\}\nonumber\\
&&+\Big([1,4'][2,1'][3,3'][4,2']+[1,2'][2,1'][3,3'][4,4']+[1,1'][2,4'][3,3'][4,2']+[1,4'][2,2'][3,3'][4,1']\Big)\{1,4'\}\{2,1'\}\{3,3'\}\{4,2'\}\nonumber\\
&&+\Big([1,2'][2,3'][3,1'][4,4']+[1,2'][2,1'][3,3'][4,4']+[1,1'][2,3'][3,2'][4,4']+[1,3'][2,2'][3,1'][4,4']\Big)\{1,2'\}\{2,3'\}\{3,1'\}\{4,4'\}\nonumber\\
&&+\Big([1,3'][2,1'][3,2'][4,4']+[1,2'][2,1'][3,3'][4,4']+[1,1'][2,3'][3,2'][4,4']+[1,3'][2,2'][3,1'][4,4']\Big)\{1,3'\}\{2,1'\}\{3,2'\}\{4,4'\}\nonumber\\
&&+4\Big([1,1'][2,2'][3,4'][4,3'] +[1,2'][2,1'][3,3'][4,4']+[1,1'][2,3'][3,2'][4,4']\nonumber\\
&&+[1,1'][2,4'][3,3'][4,2']+[1,3'][2,2'][3,1'][4,4']+[1,4'][2,2'][3,3'][4,1']\Big)\{1,1'\}\{2,2'\}\{3,3'\}\{4,4'\}\nonumber\\
&&+[1,1'][2,2'][3,3'][4,4']\Big(\{1,2'\}\{2,1'\}\{3,4'\}\{4,3'\}+\{1,3'\}\{2,4'\}\{3,1'\}\{4,2'\}+\{1,4'\}\{2,3'\}\{3,2'\}\{4,1'\}\nonumber\\
&&+\{1,1'\}\{2,3'\}\{3,4'\}\{4,2'\}+\{1,1'\}\{2,4'\}\{3,2'\}\{4,3'\}+\{1,4'\}\{2,2'\}\{3,1'\}\{4,3'\}+\{1,3'\}\{2,2'\}\{3,4'\}\{4,1'\}\nonumber\\
&&+\{1,2'\}\{2,4'\}\{3,3'\}\{4,1'\}+\{1,4'\}\{2,1'\}\{3,3'\}\{4,2'\}+\{1,2'\}\{2,3'\}\{3,1'\}\{4,4'\}+\{1,3'\}\{2,1'\}\{3,2'\}\{4,4'\}\Big)
\Bigg].\
\end{eqnarray}
Finally we need to Fourier transfer this in order to write the physical cross section. We again use the approximation of a translationally invariant target. In the Fourier space all the terms have the common factor $D(\q_1-\p_1)D(\q_2-\p_2)D(\p_3-\q_3)D(\p_4-q_4)$. 
Thus 
\begin{equation}
I_k=D(\q_1-\p_1)D(\q_2-\p_2)D(\p_3-\q_3)D(\p_4-\q_4)J_k,
\end{equation}
with
\begin{eqnarray}\label{full4p}
&&J_0=\{\p_1,\p_1\}\{\p_2,\p_2\}\{\p_3,\p_3\}\{\p_4,\p_4\},\nonumber\\
&&J_1=-\frac{1}{2N}\Bigg[
\{\p_1,\p_1\}\{\p_2,\p_2\}\{\p_3,\p_4\}\{\p_4,\p_3\}+\{\p_1,\p_1\}\{\p_2,\p_2\}\{\p_3,\p_3+\q_4-\q_3\}\{\p_4,\p_4+\q_3-\q_4\}\nonumber\\
&&+
\{\p_1,\p_2\}\{\p_2,\p_1\}\{\p_3,\p_3\}\{\p_4,\p_4\}+\{\p_1,\p_1+\q_2-\q_1\}\{\p_2,\p_2+\q_1-\q_2\}\{\p_3,\p_3\}\{\p_4,\p_4\}\nonumber\\
&&+
\{\p_1,\p_1\}\{\p_2,\p_3\}\{\p_3,\p_2\}\{\p_4,\p_4\}+\{\p_1,\p_1\}\{\p_2,\p_2+\q_3-\q_2\}\{\p_3,\p_3+\q_2-\q_3\}\{\p_4,\p_4\}
 \nonumber\\
&&+
\{\p_1,\p_1\}\{\p_2,\p_4\}\{\p_3,\p_3\}\{\p_4,\p_2\}+\{\p_1,\p_1\}\{\p_2,\p_2+\q_4-\q_2\}\{\p_3,\p_3\}\{\p_4,\p_4+\q_2-\q_4\}\nonumber\\
&&+
\{\p_1,\p_3\}\{\p_2,\p_2\}\{\p_3,\p_1\}\{\p_4,\p_4\}+\{\p_1,\p_1+\q_3-\q_1\}\{\p_2,\p_2\}\{\p_3,\p_3+\q_1-\q_3\}\{\p_4,\p_4\}\nonumber\\
&&+
\{\p_1,\p_4\}\{\p_2,\p_2\}\{\p_3,\p_3\}\{\p_4,\p_1\}+\{\p_1,\p_1+\q_4-\q_1\}\{\p_2,\p_2\}\{\p_3,\p_3\}\{\p_4,\p_4+\q_1-\q_4\}
\Bigg],\nonumber\\
&&J_2=\frac{1}{4N^2}\Bigg[
\{\p_1,\p_1+\q_2-\q_1\}\{\p_2,\p_2+\q_1-\q_2\}\{\p_3,\p_3+\q_4-\q_3\}\{\p_4,\p_4+\q_3-\q_4\}\nonumber\\
&&+\{\p_1,\p_2\}\{\p_2,\p_1\}\{\p_3,\p_3+\q_4-\q_3\}\{\p_4,\p_4+\q_3-\q_4\}+\{\p_1,\p_1+\q_2-\q_1\}\{\p_2,\p_2+\q_1-\q_2\}\{\p_3,\p_4\}\{\p_4,\p_3\}
\nonumber\\
&&+
\{\p_1,\p_3+\q_1-\q_3\}\{\p_2,\p_4+\q_2-\q_4\}\{\p_3,\p_3+\q_1-\q_3\}\{\p_4,\p_4+\q_2-\q_4\}\nonumber\\
&&+\{\p_1,\p_3\}\{\p_2,\p_2+\q_4-\q_2\}\{\p_3,\p_1\}\{\p_4,\p_4+\q_2-\q_4\}+\{\p_1,\p_1+\q_3-\q_1\}\{\p_2,\p_4\}\{\p_3,\p_3+\q_1-\q_3\}\{\p_4,\p_2\}
\nonumber\\
&&+
\{\p_1,\p_1+\q_4-\q_1\}\{\p_2,\p_2+\q_3-\q_2\}\{\p_3,\p_3+\q_2-\q_3\}\{\p_4,\p_4+\q_1-\q_4\}\nonumber\\
&&+\{\p_1,\p_4\}\{\p_2,\p_2+\q_3-\q_2\}\{\p_3,\p_3+\q_2-\q_3\}\{\p_4,\p_1\}+\{\p_1,\p_1+\q_1-\q_4\}\{\p_2,\p_3\}\{\p_3,\p_2\}\{\p_4,\p_4+\q_1-\q_4\}
\nonumber\\
&&+
\{\p_1,\p_1\}\{\p_2,\p_2+\q_3-\q_2\}\{\p_3,\p_3+\q_4-\q_3\}\{\p_4,\p_4+\q_2-\q_4\}\nonumber\\
&&
+\{\p_1,\p_1\}\{\p_2,\p_4+\q_3-\q_4\}\{\p_3,\p_3+\q_4-\q_3\}\{\p_4,\p_2\}\nonumber\\
&&+\{\p_1,\p_1\}\{\p_2,\p_2+\q_3-\q_2\}\{\p_3,\p_4\}\{\p_4,\p_3+\q_2-\q_3\}+\{\p_1,\p_1\}\{\p_2,\p_3\}\{\p_3,\p_2+\q_4-\q_2\}\{\p_4,\p_4+\q_2-\q_4\}
\nonumber\\
&&
+\{\p_1,\p_1\}\{\p_2,\p_2+\q_4-\q_2\}\{\p_3,\p_3+\q_2-\q_3\}\{\p_4,\p_4+\q_3-\q_4\}+\{\p_1,\p_1\}\{\p_2,\p_3+\q_4-\q_3\}\{\p_3,\p_2\}\{\p_4,\p_4+\q_3-\q_4\}\nonumber\\
&&+\{\p_1,\p_1\}\{\p_2,\p_4\}\{\p_3,\p_3+\q_2-\q_4\}\{\p_4,\p_2+\q_3-\q_2\}+\{\p_1,\p_1\}\{\p_2,\p_2\}+\q_4-\q_2\}\{\p_3,\p_4+\q_2-\q_4\}\{\p_4,\p_3\}\nonumber\\
&&+
\{\p_1,\p_1-\q_1+\q_4\}\{\p_2,\p_2\}\{\p_3.\p_1-\q_1+\q_3\}\{\p_4+\p_4-\q_4+\q_3\}\nonumber\\
&&+\{\p_1,\p_3-\q_3+\q_4\}\{\p_2,\p_2\}\{\p_3,\p_1\}\{\p_4,\p_3-\q_3+\q_4\}+\{\p_1,\p_4\}\{\p_2,\p_2\}\{\p_3,\p_3-\q_3+\q_1\}\{\p_4,\p_1-\q_1+\q_3\}\nonumber\\
&&+\{\p_1,\p_1-\q_1+\q_4\}\{\p_2,\p_2\}\{\p_3,\p_4-\q_4+\q_1\}\{\p_4,\p_3\}+\{\p_1,\p_1-\q_1+\q_3\}\{\p_2,\p_2\}\{\p_3,\p_3-\q_3+\q_4\}\{\p_4,\p_4-\q_4+\q_1\}\nonumber\\
&&+\{\p_1,\p_4-\q_4+\q_3\}\{\p_2,\p_2\}\{\p_3,\p_3-\q_3+\q_4\}\{\p_4,\p_1\}\nonumber\\
&&+\{\p_1,\p_1-\q_1+\q_3\}\{\p_2,\p_2\}\{\p_3,\p_4\}\{\p_4,\p_3-\q_3+\q_1\}\nonumber\\
&&+\{\p_1,\p_3\}\{\p_2,\p_2\}\{\p_3,\p_1-\q_1+\q_4\}\{\p_4,\p_4-\q_4+\q_1\}\nonumber\\
&&+\{\p_1,\p_1-\q_1+\q_2\}\{\p_2, \p_2-\q_2+\q_4\}\{\p_3,\p_3\}\{\p_4,\p_4-\q_4+\q_1\}\nonumber\\
&&+\{\p_1,\p_1-\q_1+\q_2\}\{\p_2,\p_4\}\{\p_3,\p_3\}\{\p_4,\p_2-\q_2+\q_1\}\nonumber\\
&&+\{\p_1,\p_4-\q_4+\q_2\}\{\p_2, \p_2-\q_2+\q_4\}\{\p_3, \p_3\}\{\p_4,\p_1\}+
\{\p_1,\p_2\}\{\p_2,\p_1-\q_1+\q_4\}\{\p_3,\p_3\}\{\p_4,\p_4-\q_4+\q_1\}\nonumber\\
&&+\{\p_1,\p_4-\q_4+\q_1\}\{\p_2,\p_1-\q_1+\q_2\}\{\p_3,\p_3\}\{\p_4,\p_4-\q_4+\q_2\}+\{\p_1,\p_4\}\{\p_2,\p_2-\q_2+\q_1\}\{\p_3,\p_3\}\{\p_4,\p_1-\q_1+\q_2\}\nonumber\\
&&+\{\p_1,\p_2-\q_2+\q_4\}\{\p_2,\p_1\}\{\p_3,\p_3\}\{\p_4,\p_4-\q_4+\q_2\}+\{\p_1,\p_1-\q_1+\q_4\}\{\p_2,\p_4-\q_4+\q_1\}\{\p_3,\p_3\}\{\p_4,\p_2\}\nonumber\\
&&+\{\p_1,\p_1-\q_1+\q_2\}\{\p_2,\p_2-\q_2+\q_3\}\{\p_3,\p_3-\q_3+\q_1\}\{\p_4,\p_4\}\nonumber\\
&&+\{\p_1,\p_1-\q_1+\q_2\}\{\p_2,\p_3\}\{\p_3,\p_2-\q_2+\q_1\}\{\p_4,\p_4\}+\{\p_1,\p_3-\q_3+\q_2\}\{\p_2,\p_2-\q_2+\q_3\}\{\p_3,\p_1\}\{\p_4,\p_4\}\nonumber\\
&&+\{\p_1,\p_2\}\{\p_2,\p_1-\q_1+\q_3\}\{\p_3,\p_3-\q_3+\q_1\}\{\p_4,\p_4\}\nonumber\\
&&+\{\p_1,\p_1-\q_1+\q_3\}\{\p_2,\p_2-\q_2+\q_1\}\{\p_3,\p_3-\q_3+\q_2\}\{\p_4,\p_4\}\nonumber\\
&&+\{\p_1,\p_3\}\{\p_2,\p_2-\q_2+\q_1\}\{\p_3,\p_1-\q_1+\q_2\}\{\p_4,\p_4\}+\{\p_1,\p_2-\q_2+\q_3\}\{\p_2,\p_1\}\{\p_3,\p_3-\q_3+\q_2\}\{\p_4,\p_4\}\nonumber\\
&&+\{\p_1,\p_1-\q_1+\q_3\}\{\p_2,\p_3-\q_3+\q_1\}\{\p_3,\p_2\}\{\p_4,\p_4\}
\nonumber\\
&&+4\Bigg[\{\p_1,\p_1\}\{\p_2,\p_2\}\{\p_3,\p_4-\q_4+\q_3\}\{\p_4,\p_3-\q_3+\q_4\}\nonumber\\
&&+\{\p_1,\p_2-\q_2+\q_1\}\{\p_2,\p_1-\q_1+\q_2\}\{\p_3,\p_3\}\{\p_4,\p_4\}\nonumber\\
&&+\{\p_1,\p_1\}\{\p_2,\p_3-\q_3+\q_2\}\{\p_3,\p_2-\q_2+\q_3\}\{\p_4,\p_4\}
\nonumber\\
&&+\{\p_1,\p_1\}\{\p_2,\p_4-\q_4+\q_2\}\{\p_3,\p_3\}\{\p_4,\p_2-\q_2+\q_4\}\nonumber\\
&&+\{\p_1,\p_3-\q_3+\q_1\}\{\p_2,\p_2\}\{\p_3,\p_1-\q_1+\q_3\}\{\p_4,\p_4\}\nonumber\\
&&+\{\p_1,\p_4-\q_4+\q_1\}\{\p_2,\p_2\}\{\p_3,\p_3\}\{\p_4,\p_1-\q_1+\q_4\}\Bigg]\nonumber\\
&&+\{\p_1,\p_2\}\{\p_2,\p_1\}\{\p_3,\p_4\}\{\p_4,\p_3\}+\{\p_1,\p_3\}\{\p_2,\p_4\}\{\p_3,\p_1\}\{\p_4,\p_2\}+\{\p_1,\p_4\}\{\p_2,\p_3\}\{\p_3,\p_2\}\{\p_4,\p_1\}\nonumber\\
&&+\{\p_1,\p_1\}\{\p_2,\p_3\}\{\p_3,\p_4\}\{\p_4,\p_2\}+\{\p_1,\p_1\}\{\p_2,\p_4\}\{\p_3,\p_2\}\{\p_4,\p_3\}\nonumber\\
&&+\{\p_1,\p_4\}\{\p_2,\p_2\}\{\p_3,\p_1\}\{\p_4,\p_3\}+\{\p_1,\p_3\}\{\p_2,\p_2\}\{\p_3,\p_4\}\{\p_4,\p_1\}\nonumber\\
&&+\{\p_1,\p_2\}\{\p_2,\p_4\}\{\p_3,\p_3\}\{\p_4,\p_1\}+\{\p_1,\p_4\}\{\p_2,\p_1\}\{\p_3,\p_3\}\{\p_4,\p_2\}\nonumber\\
&&+\{\p_1,\p_2\}\{\p_2,\p_3\}\{\p_3,\p_1\}\{\p_4,\p_4\}+\{\p_1,\p_3\}\{\p_2,\p_1\}\{\p_3,\p_2\}\{\p_4,\p_4\}.\
\end{eqnarray}

\begin{acknowledgments}
 A.K. would like to thank the Particles and Nuclear Physics Group of  the Universidad Santa Maria and Physics Department of the Ben Gurion University for the hospitality while this work was being done. We thank Genya Levin and Vladi Skokov for interesting and useful discussions. This research was supported in part by Fondecyt grant 1150135, ECOS-Conicyt C14E01, Anillo ACT1406 and Conicyt PIA/Basal FB0821 (A.R.) and  the NSF Nuclear Theory grant 1614640, Conicyt (MEC) grant PAI 80160015  in Chile, Fulbright US scholar program in Israel and CERN scientific associateship at CERN (A.K.).
\end{acknowledgments}


\end{document}